\documentclass[12pt]{article}

\usepackage{scicite}

\usepackage{times}
\usepackage{physunits}
\usepackage{siunitx}
\usepackage{balance}
\usepackage{graphicx}
\usepackage{media9}
\usepackage{amsmath}
\usepackage[version=4]{mhchem}
\usepackage{soul}

\newenvironment{sciabstract}{%
\begin{quote} \bf}
{\end{quote}}

\newcounter{lastnote}

\title{High-yield atmospheric water capture via bioinspired material segregation\\
\large Ultra-fast, convection-limited water capture}

\author
{Yiwei Gao$^{1}$, Santiago Ricoy$^{1}$, Addison Cobb$^{1}$,\\
Ryan Phung$^{1}$, Areianna Lewis$^{1}$, Aaron Sahm$^{1}$,\\
Nathan Ortiz$^{2}$, Sameer Rao$^{2}$ \\
and H. Jeremy Cho,$^{1\ast}$\\
\\
\normalsize{$^{1}$Department of Mechanical Engineering, University of Nevada, Las Vegas,}\\
\normalsize{4505 S Maryland Pkwy, Las Vegas, Nevada, 89154, USA}\\
\normalsize{$^{2}$Department of Mechanical Engineering, University of Utah,}\\
\normalsize{1495 E 100 S, Salt Lake City, Utah, 84112, USA}\\
\\
\normalsize{$^\ast$To whom correspondence should be addressed; E-mail:  jeremy.cho@unlv.edu.}
}

\date{}

\begin{document} 


\baselineskip24pt


\maketitle


\begin{sciabstract}
  Atmospheric water harvesting is urgently needed given increasing global water scarcity. Current sorbent-based devices that cycle between water capture and release have low harvesting rates. We envision a radically different multi-material architecture with segregated and simultaneous capture and release. This way, proven fast-release mechanisms that approach theoretical limits can be incorporated; however, no capture mechanism exists to supply liquid adequately for release. Inspired by tree frogs and airplants, our capture approach transports water through a hydrogel membrane ``skin’' into a liquid desiccant. We report an extraordinarily high capture rate of \SI{5.50}{\kilogram\per\meter\squared\per\day}at a low humidity of \SI{35}{\percent}, limited by the convection of air to the device. At higher humidities, we demonstrate up to \SI{16.9}{\kilogram\per\meter\squared\per\day}, exceeding theoretical limits for release. Simulated performance of a hypothetical one-square-meter device shows that water could be supplied to two to three people in dry environments. This work is a significant step toward providing new resources to water-scarce regions.
\end{sciabstract}

\section*{Introduction}
\begin{figure}[htb!]
    \centering
    \includegraphics[width=\textwidth]{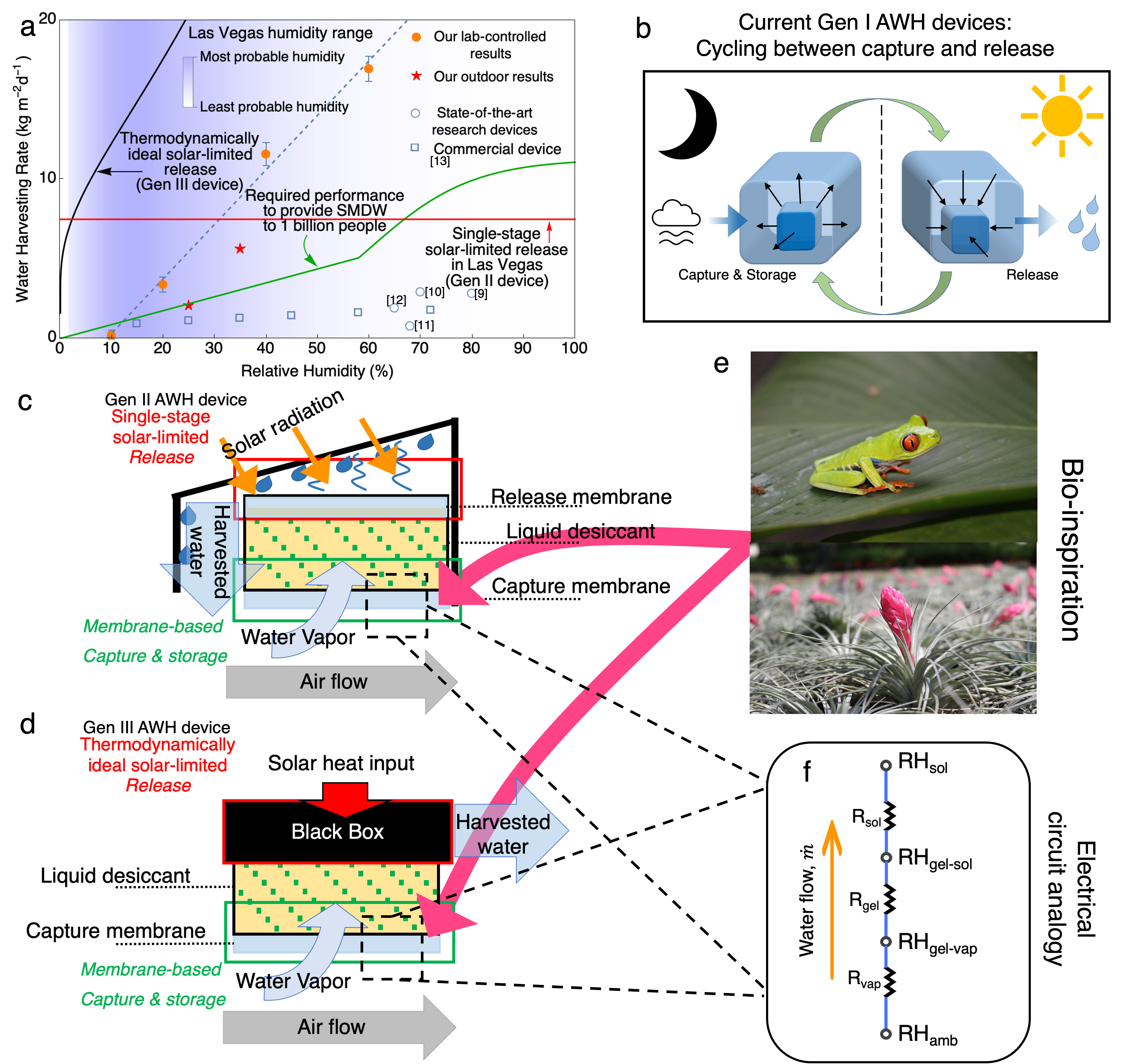}
    \caption{\textbf{Our bio-inspired water harvesting concept.} (a) None of the existing solar-based AWH state-of-the-art research \mbox{\cite{qi2019interfacial,wang2019interfacial,lapotin2021dual,ayyagari2021design}} or commercial devices \mbox{\cite{source}} meet the required performance to provide safely managed drinking water to 1 billion people as modeled by Lord et al. (green line; \mbox{Eq.~\ref{eq.j1billion}}) \mbox{\cite{lord2021global}}. (b) Current (Gen I) AWH devices use a single monolithic material that only capture/store or release water at a time. To improve AWH performance, we envision (c) Gen II and (d) Gen III AWH devices that utilize a segregated, multi-material architecture as bio-inspired by (e) tree frogs and Tillandsia ``airplants'' \mbox{\cite{tracy2011condensation,raux2020design}}. Here, we develop a hydrogel membrane for continuous and fast capture into a liquid-desiccant storage medium (bottom of (c) and (d))---a mass-transfer process we model using a (f) circuit analogy with a convective resistance in the ambient air ($R_\text{vap}$), a permeation resistance ($R_\text{gel}$) and a convective resistance in the liquid solution ($R_\text{sol}$). Coupling our capture/storage with a solar-powered release, such as a (c) single-stage distillation process \mbox{\cite{chen2019challenges,ni2016steam,tao2018solar,gao2019solar,guo2020hydrogels,shi2021all,guo2019tailoring,zhao2018highly}} or a (d) thermodynamically-limited distillation device, constitutes a complete AWH system where the harvesting rate is the minimum of capture and release rates. With some Gen II approaches already approaching the solar limit \mbox{\cite{shi2021all,guo2019tailoring,zhao2018highly}}, our fast capture/storage approach is an important component that can enable solar-limited AWH.}
    \label{fig1}
\end{figure}

Water scarcity is a serious global problem that is projected to worsen significantly in the coming decades \cite{stringer2021climate}. From a 2016 article \cite{mekonnen2016four}, around two-thirds of the global population were living under conditions of severe water scarcity for at least one month of the year. In particular, in the southwestern US, water levels are the lowest in 1,200 years \cite{williams2022rapid}. Furthermore, inadequately sanitized water sources have led to the spread of diseases such as cholera and typhoid fever, among other water-borne illnesses \cite{jarimi2020review}. Thus, technologies that can obtain clean water from alternative water resources are urgently needed. Atmospheric water harvesting (AWH) is an emerging approach to source water from the ambient air as the atmosphere contains 12.9 trillion tons of fresh water, approximately equivalent to \SI{10}{\percent} of the water in all of the lakes on Earth \cite{li2018hybrid}. AWH is especially promising to arid and landlocked regions that do not have access to surface water resources or ocean water to desalinate \cite{mekonnen2016four,kalmutzki2018metal,wang2018emerging}. These regions, like much of the southwestern US, are characterized by high temperatures, low rainfall, and low relative humidities. However, existing AWH approaches have low yields and diminishing returns at relative humidities (RH) below \SI{30}{\percent} \cite{lord2021global}. In fact, in a recent study \cite{lord2021global}, Lord et al. determined that if a particular humidity-dependent benchmark in water harvesting performance (green curve in Fig.~\ref{fig1}a, Eq.~\ref{eq.j1billion}) could be achieved, water could be provided to over one billion people in need. Yet, no existing current-generation AWH device or approach (green squares and blue triangles in Fig.~\ref{fig1}a) comes close to this benchmark.

Atmospheric water harvesting approaches can be broadly divided into two categories: fog collection and vapor condensation \cite{tu2020reviews}. The former requires humidities around \SI{100}{\percent} \cite{kalmutzki2018metal,lapotin2019adsorption,zhou2020atmospheric} and geometric structures to gather droplets from airflow or gravity \cite{klemm2012fog,andrews2011three} whereas the latter involves condensing ambient water vapor chemically using a sorbent material or thermally by cooling below the dew point \cite{eslami2018thermal,joshi2017experimental,anbarasu2011vapour}. We focus on chemical vapor condensation as this can be applied to arid, low-humidity regions where dew points can be below the freezing point, precluding thermal dew condensation techniques that are highly energy-intensive \cite{rao2022thermodynamic}. Vapor-condensation AWH broadly involves three processes: \textit{capture}, \textit{storage}, and \textit{release}. Water vapor is first \textit{captured} from the ambient environment and enters the \textit{storage} component in the harvester, which could be within a sorbent material. The subsequent \textit{release} (alternatively, distillation or desalination) stage involves evaporation (or desorption) of water from the stored component followed by condensation into fresh liquid water (single-stage distillation). This release requires substantial thermal energy \cite{rao2022thermodynamic}. As such, current-generation devices utilize solar energy for the release stage, tapping into a free, abundant and sustainable energy resource for passive operation.

However, current-generation devices produce water at rates that are far below the theoretical limits of solar-powered water release. For a single-stage distillation process (evaporation with subsequent condensation; release stage in Gen II device in \mbox{Fig.~\ref{fig1}c)}, the solar-limited water release mass flux, $j_\text{II}$, in \mbox{\si{\kilogram\per\meter\squared\per\day}}, occurs when irradiation is completely converted into latent heat:
\begin{equation}
    j_\text{II} = q_\text{solar}''/ h_\text{fg} \text{.}
    \label{eq.solar}
\end{equation}
Here, $q_\text{solar}''$ is the solar irradiation heat flux (\si{\watt\per\meter\squared}), and $h_\text{fg}$ (\si{\joule\per\kilogram}) is the specific latent heat of vaporization. Note that pure-water $h_\text{fg}$ is used for reference as this is used to defined the standard gained output ratio (GOR) in desalination devices \mbox{\cite{lienhard2017thermodynamics}}---real devices will have a slightly increased $h_\text{fg}$ per unit mass of water depending on the sorbent or desiccant used \mbox{\cite{kim2016characterization}} and as we have calculated for lithium bromide (SI Section~8). We also note that Eq.~\ref{eq.solar} is well below the second-law thermodynamic limits (we calculate the ideal device limit assuming \SI{100}{\percent} second-law efficiency in Eq.~\ref{eq.thermlimit1} and SI section~6). Nonetheless, since actual energy performance will depend on a variety of factors (including ``intermediate water" effects \mbox{\cite{zhou2019architecting}}), the pure-water $h_\text{fg}$ represents a convenient, albeit approximate, estimation of the energy usage for single-stage distillation where, at the solar-limited water flux of $j_\text{II}$, $\text{GOR}=1$. For instance, using local solar irradiation data in Las Vegas, according to Eq.~\ref{eq.solar}, $j_\text{II}$ is around \SI{7.5}{\kilogram\per\meter\squared\per\day} averaged year-round with a maximum of around \SI{12}{\kilogram\per\meter\squared\per\day} in June, which is more than sufficient to provide one's daily drinking water (\SIrange{2}{4}{\kilogram\per\day}) \cite{NAP10925} for a one-square-meter device. Furthermore, this solar limit exceeds the benchmark set by Lord et al. \mbox{\cite{lord2021global}} to reach one billion people in need for regions with humidities less than \mbox{\SI{65}{\percent}}. However, current-generation devices produce only \mbox{\SIrange{0.77}{2.89}{\kilogram\per\meter\squared\per\day}} (\mbox{Fig.~\ref{fig1}a}) \mbox{\cite{qi2019interfacial,wang2019interfacial,lapotin2021dual,ayyagari2021design}} in relatively high RH conditions of \mbox{\SIrange{65}{80}{\percent}} that are hardly representative of the arid climates that urgently require AWH. A commercial device, with performance figures at a wider range of RH conditions, produces similar fluxes at comparable conditions and lower fluxes in arid conditions \mbox{\cite{source}}. Achieving water release rates close to this solar limit, however, is not fantasy. In fact, solar distillation techniques developed within the last decade that utilize localized heating \mbox{\cite{chen2019challenges,ni2016steam,tao2018solar,gao2019solar,guo2020hydrogels,shi2021all,guo2019tailoring,zhao2018highly}} have demonstrated fluxes more than \mbox{\SI{90}{\percent}} (GOR $\sim 0.9$) of this solar limit yielding as high as \mbox{\SI{3.64}{\kilogram\per\meter\squared\per\hour}} when fully irradiated by the sun \mbox{\cite{shi2021all,guo2019tailoring,zhao2018highly}}. Pushing the limits of solar-powered release, it is possible to produce water at rates even higher than $j_\text{II}$ (GOR $>$ 1) when using multi-stage devices that approach thermodynamically ideal performance \mbox{\cite{kim2020thermodynamic, lienhard2017thermodynamics}}. This thermodynamic limit of solar-powered water release, $j_\text{III}$, quantified in \mbox{Eqs.~\ref{eq.thermlimit1}, \ref{eq.jIII}}, is an exceptionally high \mbox{\SIrange{10}{20}{\kilogram\per\meter\squared\per\day}} in typical Las Vegas conditions and much higher in more humid regions ($>\SI{100}{\kilogram\per\meter\squared\per\day}$ when $\text{RH}>\SI{75}{\percent}$).

Due to this large difference in water fluxes between realized solar-powered release devices that operate close to the solar limit \mbox{\cite{shi2021all,guo2019tailoring,zhao2018highly}} and current-generation AWH devices, we believe that current-generation devices are limited by their operational design that does not facilitate high throughput. Current-generation devices utilize a single monolithic, multifunctional sorbent material needing to be cycled between capture and release stages. This single sorbent---whether it be a metal organic framework \cite{hanikel2020mof,xu2020metal,kim2017water,ohrstrom2021improved,meng2022materials}, zeolite \cite{lapotin2021dual,lapotin2019adsorption}, hydrogel \cite{entezari2020super,zhao2019super,guo2022scalable,kallenberger2018water}, or liquid desiccant \cite{qi2019interfacial,wang2019interfacial}---has to \textit{capture}, \textit{store}, and \textit{release} water all within itself. A common benchmark of sorbents is their water \textit{uptake}, which quantifies their storage potential---a thermodynamic (equilibrium) property. Less discussed in the literature is the absorption/desorption speed of sorbents (transport/kinetics), which quantify the rates of capture and release. In any case, optimizing all of these behaviors in a single sorbent material presents significant material design and synthesis challenges. Furthermore, as this single material can only either perform capture/storage or release at any given time, \textit{it must be cycled, limiting the time window to capture as much water from the air per day} (\mbox{Fig.~\ref{fig1}b}). Finally, due to their fundamental design, incorporating existing high-flux release mechanisms relying on localized heating into single-sorbent approaches is not possible as there is no bulk liquid phase. Thus, rather than focus on improving upon existing single-sorbent approaches, we sought a complete re-design of the operation of AWH devices. In particular, we looked for ways to decouple the transport/kinetics (capture and release) from equilibrium (storage) behaviors such that they can be \textit{separately optimized to achieve higher fluxes}.

\section*{Bio-inspired design to utilize proven high-flux release mechanisms}
As opposed to the monolithic, single-material approaches, we consider a multi-material approach where the \textit{capture}, \textit{storage}, and \textit{release} are segregated into separately optimized materials. Our approach is inspired by nature where soft membranes (e.g., a tree frog's skin \cite{tracy2011condensation} or a cuticle of a \textit{Tillandsia} airplant \cite{raux2020design}; Fig.~\ref{fig1}e) can continuously capture water from the air for hydration \cite{malik2014nature}. These membranes that surround the extracellular fluid act as a protective barrier to the organism, are permeable to water, and facilitate \textit{transport} of water \cite{dominguez2011biophysical}. Meanwhile, the extracellular fluid that the membrane encases \textit{stores} the water necessary for proper hydration and survival \cite{takei2000comparative}. It also serves as a chemical potential sink creating the driving force to draw water from the air through the skin/cuticle into the fluid (\textit{capture}). As long as the chemical potential of the extracellular fluid is lower than that of the ambient water vapor, the organism will hydrate from the air. In our approach, we mimic this natural design of transport--storage segregation by having a separate transport-optimized capture membrane and a liquid desiccant to provide the chemical potential driving force and liquid storage (Fig.~\ref{fig1}c,d, green box).

Using the bio-inspired material segregation principle, we envision a vertically integrated, stacked design where a release mechanism on the top and a capture membrane on the bottom surround an interstitial storage basin of liquid desiccant (Fig.~\ref{fig1}c,d). Having the release membrane on top of a liquid desiccant (ionic solution) enables the incorporation of proven, high-yield solar release techniques that rely on heat localization at the top of a liquid phase \cite{chen2019challenges,ni2016steam,tao2018solar,gao2019solar,guo2020hydrogels,shi2021all,guo2019tailoring,zhao2018highly}. The highest-performing approaches utilize hydrogel membranes with water fluxes reported as high as \SI{3.64}{\kilogram\per\meter\squared\per\hour} under full irradiation \cite{shi2021all,guo2019tailoring,zhao2018highly}. Therefore, we are not concerned with developing 
a new release technique as many proven techniques have been conclusively demonstrated. Rather, the goal of our study is to develop a \textit{capture} and \textit{storage} approach that can meet or exceed the solar limit so as to adequately supply water for existing release techniques that can be later incorporated into a complete AWH system.

Recognizing this need, we focus on attaining the highest water fluxes through our material-segregated AWH \textit{capture} and \textit{storage} approach. In our approach, water vapor condenses and permeates through a \textit{transport}-optimized hydrogel membrane into a liquid desiccant. This desiccant is a saturated lithium bromide (LiBr) salt solution, as its equilibrium relative humidity is around \SIrange{6}{8}{\percent} for typical ambient temperatures \cite{greenspan1977humidity} (SI Section~1A). Thus, a saturated LiBr solution will be able to capture water vapor from the ambient down to this low humidity range of $<\SI{10}{\percent}$, which is lower than other strong desiccant salts such as lithium chloride and sodium hydroxide. \textit{LiBr solution also has extremely high uptake} that is comparable to the leading hydrogel-based sorbents \cite{guo2022scalable,graeber2023extreme} as we have calculated (SI Section~9; Fig.~S5). Biofouling is often a concern with membranes; however, the extreme salinity of saturated LiBr solution would most certainly prevent microbial growth \mbox{\cite{larsen1986halophilic}}. Also, lithium is a known microbe inhibitor \mbox{\cite{cox1990effect}}; thus, it is reasonable to assume that biofouling is not an issue as we apply the highest possible concentration of lithium in the liquid desiccant. Beneath this solution, we use a hydrogel membrane ``skin'' to condense and permeate water to the solution from the ambient. We emphasize that this hydrogel membrane, in vast contrast to other hydrogel-based AWH techniques \cite{graeber2023extreme,lu2022tailoring,aleid2022salting,lyu2022macroporous,shan2023all}, \textit{does not store water}---it simply acts as a transport medium. Thus, \textit{water uptake characterizations of our gel membrane are somewhat irrelevant to the water capture performance since the liquid desiccant is providing the capture driving force and storage}. In any case, we provide water absorption/uptake characterizations of our gel and liquid desiccant (Figs.~S5 and S11). We use a polyacrylamide hydrogel membrane as we are able to tune its properties to provide several benefits. The hydrogel, being permeable to the solution through its nanoporous polymer network \cite{gao2022quantifying}, serves as an extension of the liquid desiccant by bringing it in contact to the ambient air. The hydrogel is also a solid material providing protection from particulate matter with mechanical properties that we tuned to provide flexibility and strength. The high strength of the membrane allows for a large quantity of liquid desiccant to be stored above it with an extremely thin membrane (\SIrange{0.03}{0.7}{\milli\meter}) to optimize transport. Additionally, another motivating factor for using a membrane is that it acts as a physical barrier to the liquid desiccant. The membrane is porous at the polymer mesh scale of around \mbox{\SI{10}{\nano \meter}}, which can block any dust or physical contaminants and enhance the lifetime of the liquid desiccant. As shown in \mbox{Fig.~\ref{fig1}c,d}, water vapor flows from the ambient to the gel-air interface in the direction of lower vapor pressure, then through the hydrogel membrane into the solution chamber. In our envisioned complete AWH device, this solution would subsequently \textit{release} water through an existing solar-powered release mechanism with localized heating and condense it into fresh liquid water (Gen II device, \mbox{Fig.~\ref{fig1}c)}. We note that any localized heating would need to provide thermal separation from the capture membrane and the bulk of liquid desiccant solution in order to maintain high capture performance---easily achieved through thermally insulating but permeable materials (e.g, fabric insulation, foams, aerogels, etc.). Such a device \textit{would not be cycled in the typical fashion} as capture and release can be occurring independently and simultaneously.

\subsection*{Reducing mass transfer resistances}
To analyze and develop a fast capture and storage technique, we use an electrical circuit analogy to understand water transport. As shown in Fig.~\ref{fig1}f, the mass flow of water, $\dot{m}$ (\si{\kilo\gram\per\day}), is analogous to the electrical current. The driving force ``voltage" can be represented by the difference in relative humidities between the ambient and the liquid desiccant solution, $\text{RH}_\text{amb}-\text{RH}_\text{sol}$. Finally, the overall flow ``resistance" is comprised of three resistors in series: a convection resistance in the vapor phase, $R_\text{vap}$, a liquid diffusion resistance within the gel membrane, $R_\text{gel}$, and a convection resistance in the liquid solution phase $R_\text{sol}$. Thus, to maximize the rate of water capture and storage, the ``voltage" difference of $\text{RH}_\text{amb}-\text{RH}_\text{sol}$ should be maximized by ensuring the solution is as saturated with LiBr as possible to lower $\text{RH}_\text{sol}$, and the ``resistance" of $R_\text{vap}+R_\text{gel}+R_\text{sol}$ should be as small as possible. To do so, we focused our efforts to model these resistances and determine ways to minimize them.

We can directly calculate the convection resistance in the vapor resistance using the Blausius solution for flow over a flat plate \cite{blasius1907grenzschichten} (SI Section~2). The result is that the water flux is proportional to the square root of the velocity of crossflowing air, $U$, and the difference in relative humidity between the ambient and the gel--air interfacial surface, $\text{RH}_\text{amb}-\text{RH}_\text{surf}$, as described by the following circuit equation:
\begin{equation}
    j = \frac{\dot{m}}{A} =  \underbrace{\frac{0.664 D_\text{w,a}^{2/3}M_\text{\ce{H2O}} P_\text{sat}\rho_\text{air}^{1/6}}{\mu_\text{air}^{1/6} R T} \sqrt{\frac{U}{W}}}_{1/(R_\text{vap}A)}\underbrace{(\text{RH}_\text{amb} - \text{RH}_\text{surf})}_{\substack{\text{``voltage'' difference} \\ \text{between ambient and surface}}} \text{.}
    \label{eq.harvestspeed}
\end{equation}

Here, $A$ is the surface area of the gel membrane, $D_\text{w,a}$ is the binary molecular diffusion coefficient of water vapor in air, $M_\text{\ce{H2O}}$ is the molar mass of water, $P_\text{sat}$ is the saturation pressure of water, $\mu_\text{air}$ is the dynamic viscosity of air, $\rho_\text{air}$ is the density of air, $W$ is the width of the membrane along the direction of air flow (\SI{38}{\milli\meter} for our prototype), $R$ is the molar gas constant, and $T$ is the absolute temperature. The convection resistance is

\begin{equation}
     R_\text{vap} = \underbrace{\frac{\mu_\text{air}^{1/6} R T} {0.664 D_\text{w,a}^{2/3}M_\text{\ce{H2O}} P_\text{sat}\rho_\text{air}^{1/6} }}_\text{set by ambient environment} \sqrt{\frac{W}{U}} \frac{1}{A} \text{.}
     \label{eq.rvap}
\end{equation}
To minimize $R_\text{vap}$, the ratio of wind speed over width should be as large as possible. Since, from boundary layer theory, the boundary layer thickness is proportional to $\sqrt{W/U}$, the result indicates that the convection resistance is proportional to the boundary layer thickness. Therefore, placing a membrane in a windy location or using forced convection to minimize the boundary layer thickness should increase water capture and storage yield.

We note that this convection resistance would be present in \textit{any} atmospheric water harvesting device that uses vapor condensation. This is because water vapor, ultimately, is sourced from the air and would need to convect to some surface of an AWH device to be captured. Thus, the \textit{fastest} possible AWH device is one where all subsequent resistances after $R_\text{vap}$ are negligible. As a goal of our study, we seek to develop a capture membrane with a resistance, $R_\text{gel}$, that is at least an order of magnitude smaller than $R_\text{vap}$ so as to be negligible.

To develop a membrane of negligible resistance, we can first derive an expression for $R_\text{gel}$ to understand how material design parameters affect the transport behavior. Here, we build upon our previous work on the mechanical stiffness, hydraulic permeability, and relative-humidity dependencies of crosslinked hydrogels \cite{gao2021scaling,gao2022quantifying} that is based on semi-dilute polymer theory \cite{de1979scaling}. Hydrogels are a nanoporous mesh comprised of crosslinked polymer strands where the ``pores" are the spacing between the strands, which can be described as the mesh size, $\xi$. Through any porous medium, the mass flow of water is dictated by Darcy's law:
\begin{equation}
    j = -\rho_\text{l}\frac{\kappa}{\mu_\text{l}} \nabla P \text{,}
    \label{eq.darcy}
\end{equation}
where $\rho_\text{l}$ is the density of liquid water, $\kappa$ is the hydraulic permeability, $\mu_\text{l}$ is the dynamic viscosity of liquid water, and $\nabla P$ is pressure gradient ($\approx -\Delta P/L$ where $L$ is the thickness of the membrane). As hydrogels are poroelastic materials, the stiffness of a hydrogel (bulk modulus), $K$, is related to the changes in pressure and volume such that $K = V dP/dV$ \cite{gao2021scaling}. Defining a filling fraction, $s\equiv V/V_\text{wet}$, as the volume of a hydrogel over its wet-state, \SI{100}{\percent}-RH volume, we can express the change in pressure as
\begin{equation}
    dP = \frac{K}{V}dV = \frac{K}{s}ds \text{.}
\end{equation}
Here, $s$ can approach zero for highly de-swollen gels and be equal to unity when equilibrated in pure water. In terms of the this filling fraction, Darcy's law can be expressed as
\begin{equation}
    j = -\rho_\text{l} \frac{\kappa K} {\mu_\text{l}s} \nabla s \text{,}
    \label{eq.dpe}
\end{equation}
where $\nabla s \approx -\Delta s/L$ is the gradient in filling fraction such that water moves from regions of high filling fraction to low filling fraction. We also note that $\kappa K/\mu_\text{l}$ is known as the poroelastic diffusion coefficient  \cite{louf2021under,louf2021poroelastic}, which describes the diffusion rate of water through poroelastic media. In our previous study \cite{gao2021scaling}, we established that the filling fraction is a function of relative humidity, $s(\text{RH})$ (related to the water uptake isotherm), which can be experimentally measured using a vapor sorption analyzer (SI Section~5, Fig.~S11). Expressing the gradient of filling fraction in terms of the dependence on humidity results in the following circuit equation for water flow ``current":
\begin{equation}
    j = \frac{\dot{m}}{A} = \underbrace{\rho_\text{l} \frac{\kappa K}{\mu_\text{l} s L} \frac{ds}{d\text{RH}}}_{1/(R_\text{gel}A)} \underbrace{(\text{RH}_\text{surf}-\text{RH}_\text{sol})}_{\substack{\text{``voltage'' difference} \\ \text{between surface and solution}}} \text{.}
    \label{eq.mgel}
\end{equation}
Thus, the corresponding resistance within the gel membrane, $R_\text{gel}$, is 
\begin{equation}
    R_\text{gel} = \frac{\mu_\text{l} s L}{\rho_\text{l} \kappa K A} \frac{d\text{RH}}{ds} \text{.}
    \label{eq.rgeli}
\end{equation}
This equation, however, relies on properties that change depending on the humidity. That is, stiffness, $K$, permeability, $\kappa$, and thickness, $L$, would change with filling fraction, $s(\text{RH})$, which is a function of humidity. As we investigated in detail previously \cite{gao2022quantifying}, the permeability scales as $\kappa = \kappa_\text{wet} s^2$, where $\kappa_\text{wet}$ is the experimentally measurable wet-state value. In addition, the stiffness scales as $K = K_\text{wet}s^{-9/4}$, where $K_\text{wet}$ is the experimentally measurable wet-state bulk modulus \cite{gao2021scaling}. Finally, it can be shown that for a de-swollen hydrogel with a Poisson ratio of $1/3$ \cite{wyss2010capillary,Geissler1981,Andrei1998}, the thickness of the gel membrane when constrained to constant area, $A$, is $L=L_\text{wet}s^{4/3}$ (SI Section~3, SI Fig.~S6). All other parameters do not change with the humidity. Thus, incorporating all of the humidity dependencies into our expression of gel resistance Eq.~\ref{eq.rgeli},
\begin{equation}
    R_\text{gel} = \frac{\mu_\text{l}}{\kappa_\text{wet}K_\text{wet}}\frac{L_\text{wet}}{\rho_\text{l}  A} s^{31/12} \frac{d\text{RH}}{ds}.
    \label{eq.rgel}
\end{equation}
From Eq.~\ref{eq.rgel}, we can see that the wet-state permeability and stiffness should be maximized while the thickness should be minimized. Furthermore, highly de-swellable gels, such that $s$ is very small at operating conditions, would also minimize $R_\text{gel}$ due to the $s^{31/12}$ factor. Note that in our expression for $R_\text{gel}$, we are assuming uniform properties within the gel membrane. This is valid as long as a poroelastic Peclet number is less than unity, which we justify in SI Section~4. We also reemphasize that the gel membrane \textit{does not act as a storage medium; thus, water uptake of the gel itself is not important in determining the storage performance}---instead, storage is provided by the desiccant solution, which has similar uptake to the leading hydrogel-based sorbents \cite{guo2022scalable,graeber2023extreme} (SI Section~9; Fig.~S5). Here, we use the filling fraction $s$ as a convenient gel property to describe its thermodynamic state and \textit{not as a water capture performance metric}.

Based on our analysis of mass-transfer resistances of the gel, we sought to synthesize hydrogels with very strong type-II or type-III isotherm behavior in order to minimize the filling fraction, $s$. Thus, we used polyacrylmaide hydrogels as they are known to have very strong type-II isotherm behavior \cite{mittal2020adsorption} with small $s$ at low RH as we experimentally verified previously \cite{gao2021scaling} and for the current work (SI Fig.~S11). To create thin, \SI{0.03}{\milli\meter} gels, we strengthened the hydrogel using insights gained from a recent study on highly entangled polymer networks with minimal crosslinker \cite{kim2021fracture} (SI Section~1B). According to their analysis, hydrogels in which entanglement greatly outnumbers crosslinking have significantly higher toughness, strength, and fatigue resistance, compared to traditional crosslinking-dominant hydrogels.

Our gels had a wet-state bulk modulus, $K_\text{wet}= 27.6\pm 0.2$ $\si{\kilo\pascal}$ and a maximum strain of $\sim \SIrange{160}{200}{\percent}$. The thickness of the gel membranes, when constrained to a fixed area and subjected to the saturated LiBr environment is around $L=\SI{0.03}{\milli\meter}$. These membranes, when supported by a metal mesh, can withstand the compressive stress associated with $\sim \SI{10}{\centi\meter}$ liquid desiccant above it, equivalent to $\rho  g  h = \SI{1.4}{\kilo \pascal}$, which is much smaller than the compressive strength of the material.  The hydraulic permeability of the gels were measured using a custom flow cell \cite{gao2022quantifying} to be $\kappa_\text{wet}=\SI{7.2e-18}{\meter\squared}$. With full experimental characterization of our synthesized gels, we were able to directly calculate the gel resistance to be $R_\text{gel}=\SI{0.21e6}{\second\per\kilogram}$. Compared to the vapor resistance of $R_\text{vap}=\SI{1.84e6}{\second\per\kilogram}$, assuming an ambient temperature of \SI{23}{\celsius} and cross-flow velocity of \SI{0.3}{\meter\per\second}, $R_\text{gel}$ was approximately an order of magnitude lower than $R_\text{vapor}$. Thus, we expect our gel membranes to have negligible resistance. Furthermore, we calculated the resistance in the solution phase to be $R_\text{sol} = \SI{0.093e6}{\second\per\kilogram}$. This resistance is low due to the Rayleigh-B\'enard mixing that occurs from lower-density, lower-salt-concentration liquid at the gel--solution interface (a calculation of $R_\text{sol}$ is in SI Section~7). Since $R_\text{sol}$ is lower than $R_\text{vap}$ by at least one order of magnitude, it is also negligible. Therefore, the mass transfer in the capture stage should be vapor-convection-limited---neither gel-diffusion-limited nor solution-convection-limited---where $\dot{m}\approx(\text{RH}_\text{amb}-\text{RH}_\text{sol})/R_\text{vap}$ since $R_\text{gel} \ll R_\text{vap}$. Modeled convection-limited water capture fluxes at $\approx \SI{23}{\celsius}$ (lab conditions) at different air velocities and humidities are shown in SI Fig.~S7.

\section*{Results}
\subsection*{Lab testing convection-limited capture (indoor)}
\begin{figure}[htb!]
    \centering
    \includegraphics[width=\textwidth]{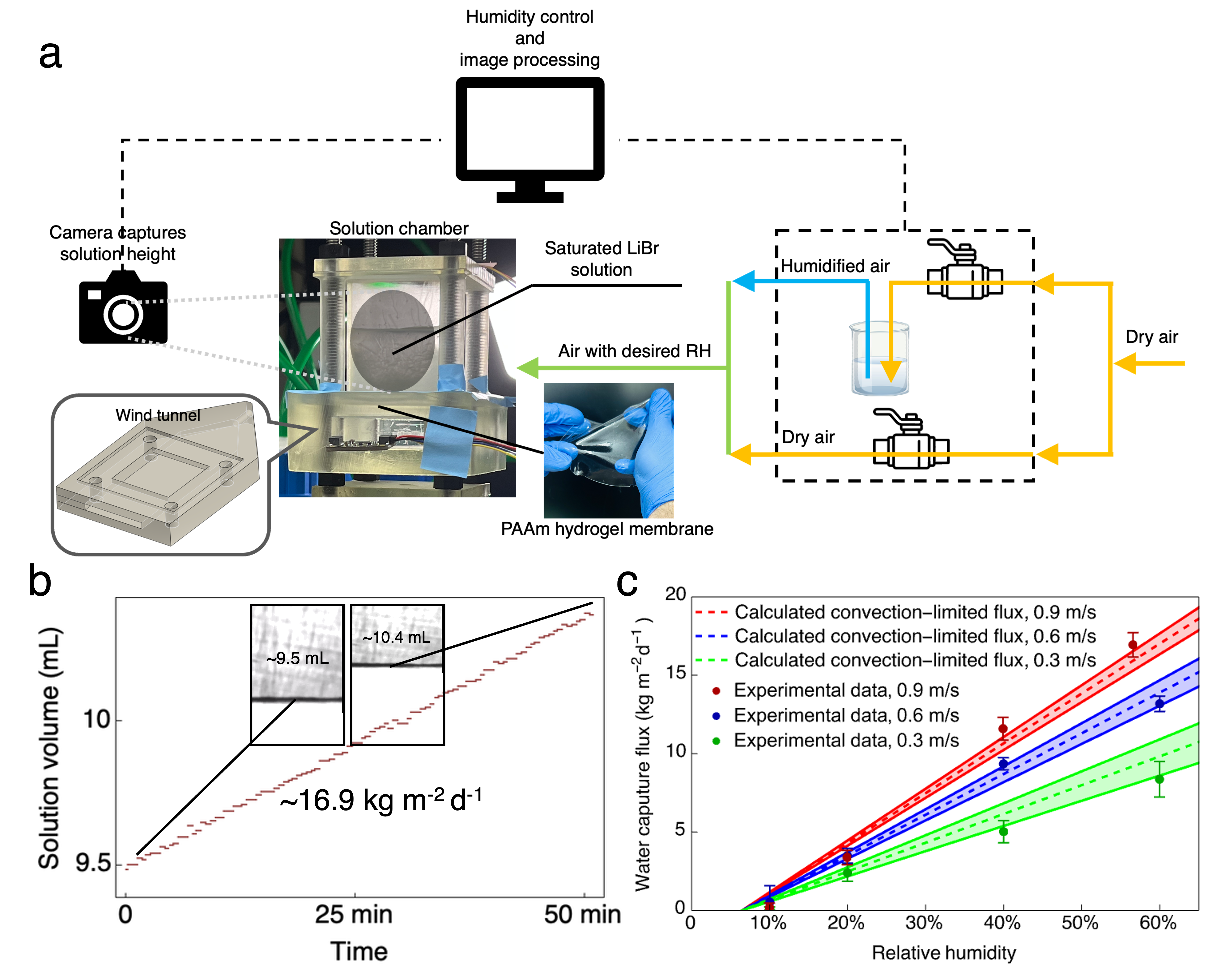}
    \caption{\textbf{Lab-controlled water capture tests confirm near-convection-limited performance.} (a) A scheme of indoor capture testing. Dry air is first humidified to a desired RH level using PID control. Humidified air with \SIrange{10}{60}{\percent} RH flows into the 3D-printed wind tunnel, located underneath a gel membrane in contact with the liquid desiccant solution above it. (b) A clear volume change of the solution at a steady rate (\SI{57}{\percent}, \SI{0.9}{\meter\per\second}, \SI{75}{min}) is shown (Supplementary Video). (c) Across 12 individual indoor vapor \textit{capture} tests with varied humidities and wind speeds, all data points collapse along predicted capture rates (dashed lines) within the range of uncertainties in velocity ($\pm \SI{0.07}{\meter\per\second}$ (1 SD); semi-transparent areas). Error bars represent 1 SD. The results confirm that the water capture and storage system is operating with convection-limited behavior and that mass transfer resistance through the gel can be neglected.}
    \label{fig2}
\end{figure}

To test our expectation of convection-limited behavior, we performed $12$ independent indoor capture/storage tests under varied conditions with crossflowing air speeds from \SIrange{0.3}{0.9}{\meter\per\second} and humidities from \SIrange{10}{60}{\percent} (SI Section~1D; plotted experimental results in Fig.~\ref{fig2}c). To facilitate analysis and comparison with established fluid dynamic models, we designed a wind tunnel to provide laminar flow (SI Section~1C). With negligible gel resistance, the mass flow rate is $\dot{m}\approx(\text{RH}_\text{amb}-\text{RH}_\text{sol})/R_\text{vap}$, where $R_\text{vap}$ can be determined ab initio to predict flow rate (dotted lines in Fig.~\ref{fig2}c) according to Eq.~\ref{eq.harvestspeed}, where $\text{RH}_\text{surf}=\text{RH}_\text{sol}$. The change of the liquid desiccant volume, $\Delta V$, was determined by photography and image processing. This volume change was converted into captured water mass, $m_\text{capture}$ (SI Section~1E). Comparing experimental capture rates with predicted rates from Blausius' exact solution for convective flow (SI Section~2), we found a remarkably close agreement between the results with \textit{no fitting}, confirming convection-limited behavior. As illustrated in the plots, water capture rate increased linearly with ambient relative humidity. Additionally, the rate increased with the square root of wind speed, $\sqrt{U}$, as expected from Eq.~\ref{eq.harvestspeed}. Further enhancement could be provided by turbulent air flow as the boundary layer would become very thin.

\subsection*{Outdoor water capture results (outdoor)}
\begin{figure}[htb!]
    \centering
    \includegraphics[width=\textwidth]{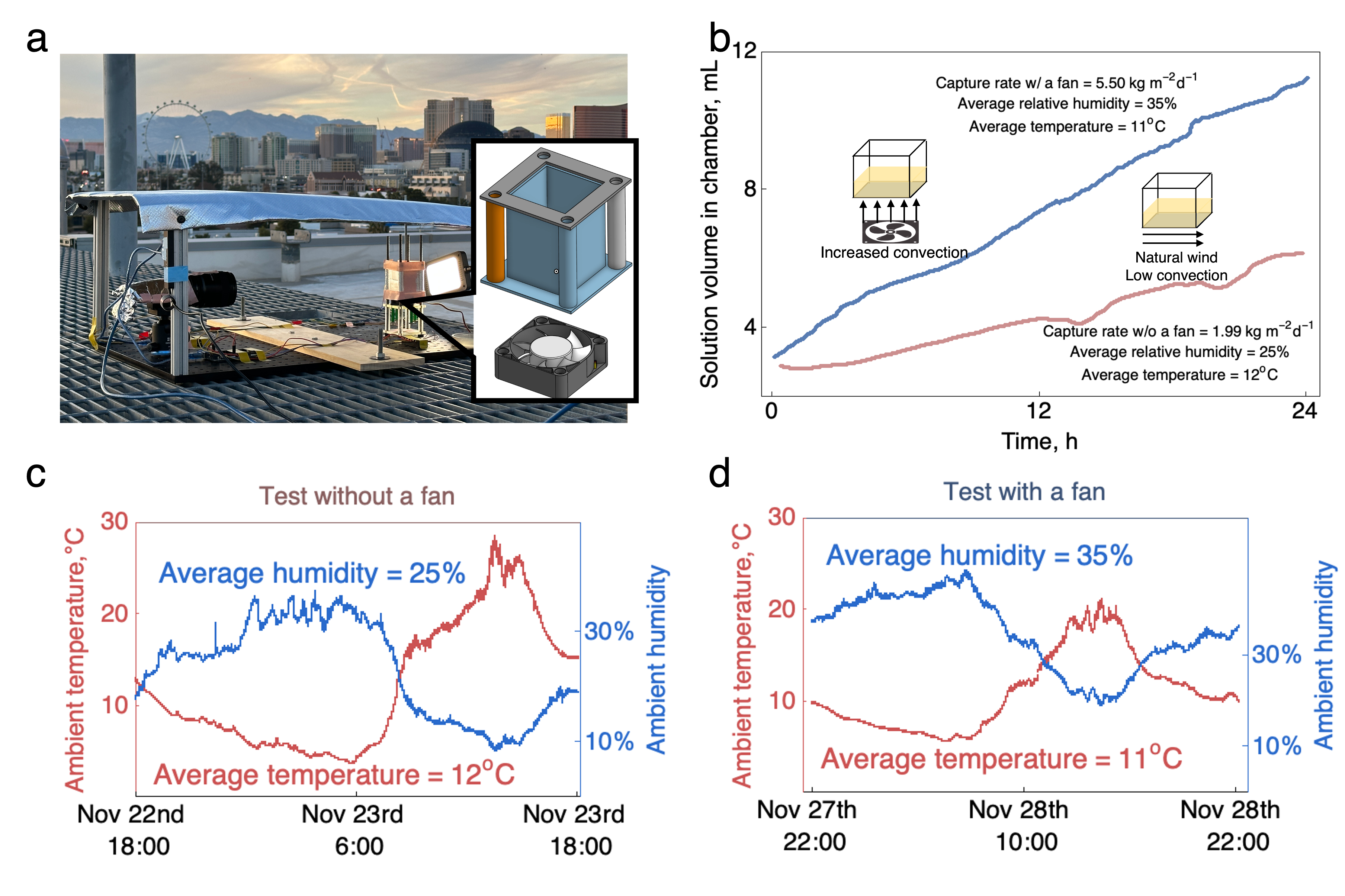}
    \caption{\textbf{Outdoor water capture tests demonstrated high water capture in Las Vegas---the driest city in the United States.} (a) Outdoor capture testing with a \SI{50}{mm} diameter fan resulted in (b) \SI{5.50}{\kilogram\per\meter\squared\per\day} over a \SI{24}{h} period in late November of 2022. Only relying on natural wind without a fan resulted in \SI{1.99}{\kilogram\per\meter\squared\per\day}. Throughout the testing periods, temperature and humidity varied as shown in (c) and (d).}
    \label{fig3}
\end{figure}

To demonstrate the potential of our AWH approach in an arid environment, we performed \textit{outdoor} capture tests locally in Las Vegas (lowest-rainfall metropolitan area in the US), where the ambient humidity ranges from \SIrange{20}{40}{\percent} in late November. Each outdoor test ran for at least \SI{24}{\hour} continuously, and both ambient temperature and humidity changes were recorded as shown in Fig.~\ref{fig3}. We compared our sensor reading on temperature and humidity and the weather data from the nearby KLAS airport obtained from Wolfram Research \cite{wolframweather}, and confirmed that our measurements of the ambient environment during outdoor tests were reliable (SI Fig.~S9). Relying only on natural wind to flow across the gel membrane, we measured an average capture rate of \SI{1.99}{\kilogram\per\meter\squared\per\day} at an average RH of \SI{25}{\percent} over a 24-h period (Fig.~\ref{fig3}b, red). We note that positive water capture was recorded even as the humidity dipped below \SI{10}{\percent} (Fig.~\ref{fig3}c), demonstrating water capture at the lowest relative humidities compared to other approaches (Fig.~\ref{fig1}c). Based on the insight gained from Eq.~\ref{eq.rvap}, where mass transfer is inversely proportional to the boundary layer thickness, we incorporated a small \SI{1.4}{\watt} fan to apply forced convection over the membrane (Fig.~\ref{fig3}a). The water capture and storage yield increased to \SI{5.50}{\kilogram\per\meter\squared\per\day} at \SI{35}{\percent}, representing the highest water yield for any AWH approach at any humidity. We note that the purpose of adding a fan was to demonstrate convection-limited transport behavior and not a technological feature.

\begin{figure}[htb!]
    \centering
    \includegraphics[width=0.95\textwidth]{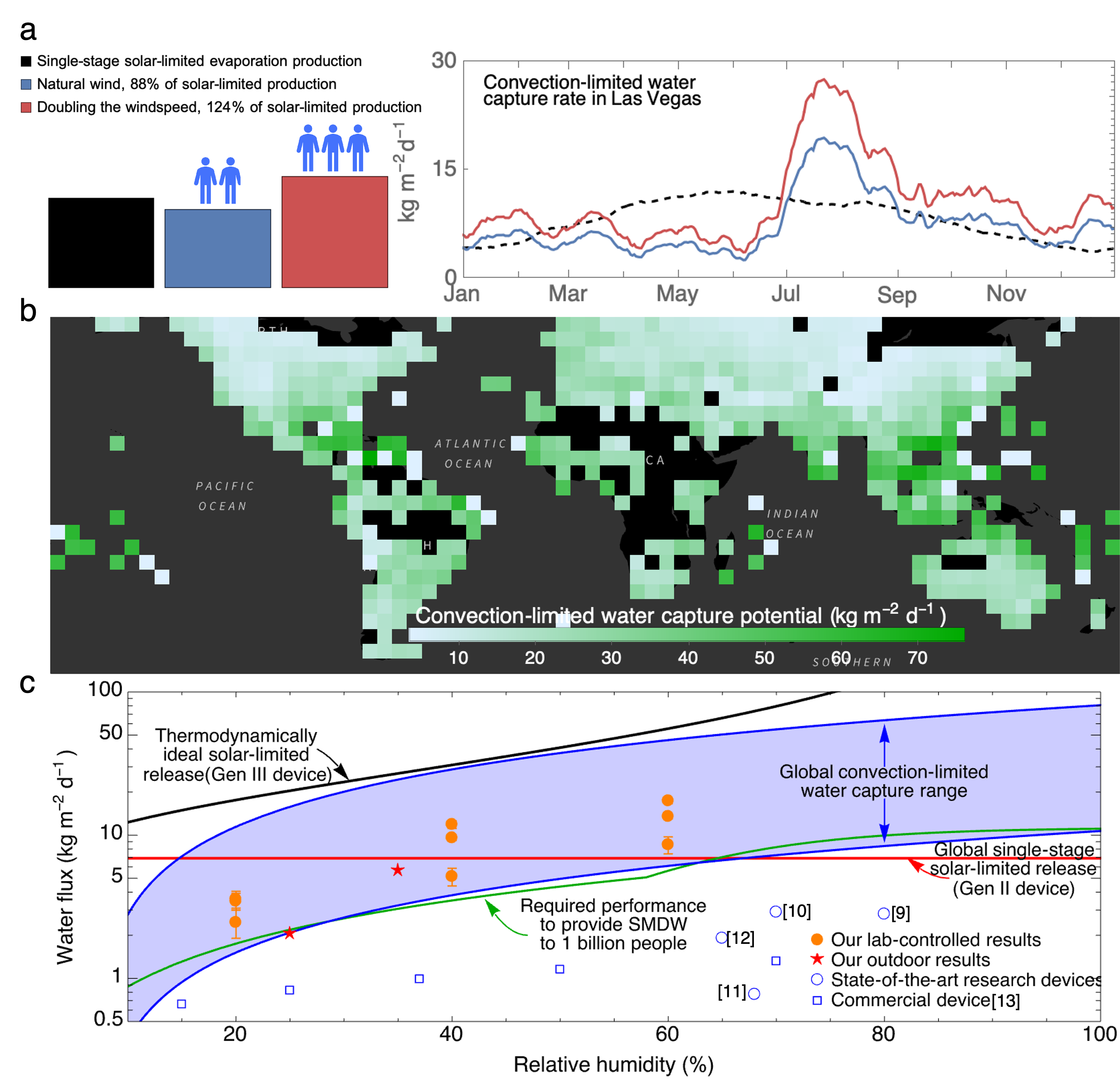}
    \caption{\textbf{Global implications of convection-limited water capture.} (a) Simulated year-round water capture rates of a convection-limited water harvesting device with the same width as our prototype ($W=\SI{38}{\milli\meter}$) in Las Vegas relying on natural wind (blue) is within \SI{88}{\percent} of the solar limit (black). Doubling the wind speed (red), (e.g., forced convection) can result in capture fluxes that exceed the solar limit. Thus, a hypothetical one-square-meter device could provide two to three adults' daily drinking water. Plotted weather data from KLAS airport obtained from Wolfram Research \cite{wolframweather} is shown in Fig.~S9. (b) We simulate convection-limited water capture potential globally and find that certain regions can exceed \SI{70}{\kilogram\per\meter\squared\per\day} (dark green regions). (c) Lord et al. \cite{lord2021global} presented a benchmark curve (green; Eq.~\ref{eq.j1billion}) for the required performance of an AWH device to provide safely managed drinking water (SMDW) to one billion people. Relying on natural wind at 1 m height, convection-limited water capture with our prototype width results in a range of fluxes (blue shaded region representing \SI{95}{\percent} of the global flux range) that generally exceed the performance required by Lord's curve. Error bars represent 1 SD.}
    \label{fig4}
\end{figure}

\section*{Further implications}
Both lab and outdoor test results confirm that, through our bio-inspired design, water can be captured at relative humidities as low as $\sim\SI{10}{\percent}$ and at rates that are close to the solar limit of water release at higher humidities. Faster rates can be achieved at higher humidities, and we have recorded as high as \SI{16.9}{\kilogram\per\meter\squared\per\day} at \SI{57}{\percent} RH. Faster rates can also be achieved with more convection. Using weather data for 2021 in Las Vegas, we modeled the performance of our AWH capture device using our experimentally verified Eq.~\ref{eq.harvestspeed}. As shown in Fig.~\ref{fig4}a, we compared simulated year-round capture rates of our AWH capture device with and without a fan for 2021 (blue and red, respectively). Relying only on natural wind (current setup), we modeled a capture rate of \SI{6.6}{\kilogram\per\meter\squared\per\day} (annual average), which is \SI{88}{\percent} of the solar limit of \SI{7.5}{\kilogram\per\meter\squared\per\day}. Doubling the wind speed either by forced convection or locating the device in a high-wind area (e.g., higher elevations or at constrictions between buildings), we modeled a capture rate of \SI{9.3}{\kilogram\per\meter\squared\per\day}, which is \SI{124}{\percent} of the solar limit. In any case, coupling our capture approach with a proven release technique would enable a complete AWH package with near-solar-limit performance. A hypothetical one-square-meter device, with a $W=\SI{38}{\milli\meter}$ width (same as our prototype) could provide individual drinking water (\SI{3}{\kilogram \per \day} \cite{NAP10925}) for two to three people in Las Vegas, the driest city in the United States. We note that any future scale up of this device would need to incorporate a release mechanism and consider appropriate mechanical frame design to distribute hydraulic stresses over a large area of gel membrane.
 
Assuming an appropriate release mechanism is incorporated with our capture and storage approach to form a complete AWH device, we simulated the global convection-limited water capture potential of our design based on global weather data sets of temperature, humidity, and wind speed \cite{gwa,hadisdh} (SI Figs.~S15--S19) as shown in Fig.~\ref{fig4}b. In nearly all land regions, the water flux is greater than \SI{10}{\kilogram\per\meter\squared\per\day}. Simulating the performance of our device across the globe (Fig.~\ref{fig4}c, blue region) results in the range of water fluxes that generally well exceed Lord's required performance curve to provide safely managed drinking water to $1$ billion people \cite{lord2021global} (Fig.~\ref{fig4}c, green line). In fact, the convection-limited fluxes can exceed the single-stage solar limit and even approach the thermodynamic solar limit for a distillation process \cite{lienhard2017thermodynamics} (SI Section~6). Furthermore, our AWH approach has potentially substantial economic advantages---we can estimate a cost of capture and storage materials to be approximately \$\SI{17}{\per\meter\squared} or \$\SI{1}{\per\kilo\gram}---considering only the cost of raw materials purchased at bulk scales, and the combined dry mass of LiBr salt and hydrogel membrane. Thus, high-yield, convection-limited atmospheric water harvesting is highly feasible and could potentially be developed at low cost for wide-scale implementation if the release mechanism could be similarly affordable. Future work will focus on complications that could arise from scaling up (e.g., turbulent wind flows, large membrane fabrication). Future studies will also investigate how a release mechanism could be incorporated to realize a complete AWH system.

 In summary, we have reimagined the atmospheric water harvesting process and envisioned a new, multilayer architecture resembling the function of skins and cuticles in nature. With this architecture, we are able to segregate the capture, storage, and release of water into separately optimized materials. The architecture also accommodates proven, highly effective water release techniques with near-solar-limit performance for single-stage distillation \cite{shi2021all,guo2019tailoring,zhao2018highly}. To supply adequate water to the release stage, we have focused on the capture and storage stages and developed a hydrogel membrane ``skin" coupled with a liquid desiccant. Using detailed transport and material analysis, we designed the membrane to provide the fastest possible water capture rates as limited by the supply of ambient air flow to the device. This is possible through the use of high entangled polymer networks for high strength, allowing gels to be made extremely thin. Detailed lab and outdoor testing demonstrate that our device has the highest capture fluxes and the largest operational humidities compared to the state of the art. We have modeled the global impact of this convection-limited performance and have found that a hypothetical one-square-meter device could provide daily water needs to several individuals in even the driest environments. Using criteria developed from a previous analysis by Lord et al.\cite{lord2021global}, implementing a device with convection-limited performance in regions without safely managed drinking water could provide water to over a billion people. Our work could be an important step toward building a scalable and affordable AWH device that can provide additional water security to arid communities with dwindling water supplies or communities with limited infrastructure.

\section*{Methods}

\subsection*{Synthesis of hydrogels}
The mixtures of acrylamide monomers, photoinitiator (Irgacure 2959) and crosslinker (N,N'-methylene(bis)acrylamide) were first made as a solution and poured into a mold with \SI{0.5}{\milli\meter} thickness. UV irradiation was applied for \SI{1}{\hour}. The crosslinked hydrogel samples were carefully removed from the mold and rinsed to remove unreacted chemicals. All samples were immersed in DI water for \SI{3}{\day} until reaching the equilibrium wet state before use.

\subsection*{Saturated salt solution}
From Greenspan \cite{greenspan1977humidity} (SI Section~1A), the equilibrium relative humidity of saturated lithium bromide (LiBr) solution is $\approx$ \SI{8}{\percent} at room temperature. To prepare LiBr saturated solutions, LiBr salt was gradually added into DI water and mixed by a magnetic stirrer, until a solid phase was precipitated. The liquid was allowed to cool after natural exothermic heating from dissolution.

\subsection*{Tensile testing}
We used a custom-built tensile/compression tester used previously \cite{gao2021scaling} to stretch six dogbone-shape hydrogel samples. From stress-strain curves (SI Fig.~S13), we determined the bulk modulus, $K_\text{wet}$, from the Young's modulus assuming Poisson's ratio is $1/3$ \cite{wyss2010capillary,Geissler1981,Andrei1998}. To ensure the hydrogel samples were tested at their wet state, all tests were finished within \SI{5}{\minute} of removal from water.

\subsection*{Permeability testing}
We measured the hydraulic permeability of hydrogel membranes, $\kappa_\text{wet}$, with a custom-built permeability tester used previously \cite{gao2022quantifying} (SI Fig.~S14). In the previous work, we found that the volumetric flow rate, $Q$, was linear with $\Delta P$ within $\pm \SI{2}{\percent}$ when $\Delta P/K_\text{wet}$ was in the range of \SI{0.5} - \SI{1}; thus, we applied a pressure, $\Delta P = \SI{70}{\percent} K_\text{wet}$, by the Elveflow Microfluidic Flow Controller and recorded the real-time water volumetric flow rate for \SI{30}{\minute} to calculate the hydraulic permeability of the sample based on Darcy's law. The scheme of the custom-built permeability tester is shown in SI Fig.~S14.

\subsection*{Indoor capture and storage testing}
We used a custom-built wind tunnel with PID control of humidity in the range of \SIrange{10}{60}{\percent} and varied mean velocities from \SIrange{0.3}{0.9}{\meter\per\second}. Air flow consideration in the wind tunnel is shown in SI Section~1C. A camera captured the height of liquid meniscus every \SI{30}{\second}, enabling us to determine the change of saturated solution, $V_\text{change}$, in the chamber and calculate the water capture rate over time. The solution humidity was monitored to ensure it remained at $\approx \SI{8}{\percent}$ during the entirety of the tests. Further details of indoor testing procedures are shown in SI Section~1D and SI Fig.~S8. 

\subsection*{Calculation of captured water mass}
As the LiBr solution in the chamber was maintained at saturation, the volume change of the liquid, $\Delta V$, came from both the volume change of saturated solution that consisted of captured water and dissolved salt, and the volume change of undissolved salt. The mass of captured water, $m_\text{capture}$, can be calculated as
\begin{equation}
    m_\text{capture} = \Delta V \frac{\left(1-w\right) \rho_\text{salt}\rho_\text{solution}}{\rho_\text{salt}-w\rho_\text{solution}},
\end{equation}
where $w$ is the mass fraction of LiBr in water at the solubility limit, $\rho_\text{salt}$ is the density of LiBr salt, and $\rho_\text{solution}$ is the density of LiBr saturated solution as a function of temperature. Detailed discussion and derivation are shown in SI Section~1E.

\subsection*{Outdoor capture and storage tests}
The outdoor test setup was identical to the indoor setup with the absence of the wind tunnel and humidity control system. We tested with and without a \SI{50}{\milli\meter} computer fan that consumed \SI{1.4}{\watt} of electrical power during operation. Each outdoor test was operated on the roof of a laboratory building for at least \SI{24}{\hour} continuously. Ambient humidity and temperature were measured and recorded during the entire experiment process for analysis. 

\subsection*{Water sorption testing}
We applied dynamic vapor sorption (DVS) to determine the sorption response of the hydrogel in varied humidities with the DVS Adventure from Surface (SI Fig.~S10). The sample was exposed to progressively lower RH conditions in \SI{10}{\percent} decrements, with smaller changes when near the saturation point, and allowed to equilibrate at each condition from \SIrange{98}{10}{\percent}. The mass fraction isotherm was fit to a GAB isotherm model \cite{mittal2020adsorption} (SI Fig.~S11). Further details of water sorption testing and modeling are shown in SI Section~5 and SI Figs.~S10 \& S11.

\subsection*{Lord et al.'s harvesting performance benchmark}
Lord et al. \cite{lord2021global} determined the required \textit{specific yield} of water to supply one billion people with safely managed drinking water (SMDW) taking into account global data on local population distribution, local solar irradiance, local humidity, and local water need. Humidity-dependent specific yield was expressed in \si{\kilogram} of water produced per \si{\kilo\watt\hour} of solar energy. In our work, we use quantify capture or harvesting performance as a mass flux in \si{\kilogram\per\square\meter\per\day}; thus, to convert specific yield to a mass flux, we multiply the specific yield by the global horizontal irradiance (GHI), equivalent to the incoming solar radiation on a flat surface per unit area. In Figs.~\ref{fig1} and \ref{fig4} (green curve), we take the maximum of the two curves presented by Lord et al., linear and logistic where
\begin{equation}
\begin{split}
    j_\text{1billion}=\text{max}\left(\underbrace{\left(\SI{1.86}{\kilogram\per\kilo\watt\per\hour}\right) \text{RH}_\text{amb}}_\text{Lord's linear curve}, \underbrace{\frac{\SI{2.4}{\kilogram\per\kilo\watt\per\hour}}{1+\exp\left(-10.0\left(\text{RH}_\text{amb}-0.6\right)\right)}}_\text{Lord's logistic curve}\right) \\
    \times q''_\text{solar}
\label{eq.j1billion}
\end{split}
\end{equation}
and $q''_\text{solar}$ is global-average GHI of \SI{4.7}{\kilo\watt\hour\per\square\meter\per\day}. Thus, we are showing a conservative mass flux requirement to supply one billion people with SMDW.

\subsection*{Modeling location-specific water capture potential}
Global solar-limited water release was determined using Eq.~\ref{eq.solar} using global horizontal irradiance data from the Global Solar Atlas 2.0 \cite{gsa}. Convection-limited water capture flux potential was calculated using Eq.~\ref{eq.harvestspeed}, where $\text{RH}_\text{surf} = \text{RH}_\text{amb}$, assuming $R_\text{gel} \ll R_\text{vap}$. Values for the diffusion coefficient, $D_\text{w,a}$, were determined using values and an equation from \mbox{\cite{incropera1996fundamentals}}. Water and humid air properties were determined using CoolProp \mbox{\cite{coolprop}}. Local wind speed data were taken from Wolfram Research \mbox{\cite{wolframweather}}, while global wind speed data were taken from the Global Wind Atlas \mbox{\cite{gwa}}. 10-m wind speeds were converted to 1-m wind speeds using the power-law wind profile with an exponent of $1/7$ \mbox{\cite{touma1977dependence}}. Global humidity and temperature data were taken from the HadISDH.blend 1.3.0.2021f version of the Met Office Hadley Centre Integrated Surface Dataset of Humidity \mbox{\cite{hadisdh}}.

\subsection*{Solar limit with ideal distillation}
For a single-stage distillation system with no heat recovery, the energy required to distill water is the latent heat, as in Eq.~\ref{eq.solar}. However, in a thermodynamically reversible (100\% second-law efficiency; no entropy generation) black box with an inflow of saturated salt solution, the solar heat required, $Q_\text{h}$, to produce a flow of distilled water, $\dot{m}_\text{water}$, is
\begin{equation}
    \frac{Q_\text{h}}{\dot{m}_\text{water}}=\frac{R T_\text{amb} \ln\left(\frac{1}{\text{RH}_\text{sat}}\right)}{M_\text{water}\left(1 - \frac{T_\text{amb}}{T_\text{h}}\right)} {,}
    \label{eq.thermlimit1}
\end{equation}
where $R$ is the molar gas constant, $T_\text{amb}$ is the ambient temperature, $\text{RH}_\text{sat}$ is the equilibrium relative humidity of the saturated salt solution, $M_\text{water}$ is the molar mass of water, and $T_\text{h}$ is the temperature of the heat source (assumed to be at the boiling point of water at \SI{373}{K} in Fig.~\ref{fig4}c). Note that this device inherently utilizes a Carnot heat engine to produce work: $W=Q_\text{h}\left(1- T_\text{amb}/T_\text{h}\right)$. For a solar-powered release system, where $Q_\text{h} = Q_\text{solar}$, the thermodynamically limited mass flux is

\begin{equation}
    j_\text{III} = \frac{\dot{m}_\text{water}}{A} = q''_\text{solar}\left(1-\frac{T_\text{amb}}{T_\text{h}}\right)\frac{M_\text{water}}{R T_\text{amb} \ln \left(\frac{1}{\text{RH}_\text{sat}}\right)}{,}
    \label{eq.jIII}
\end{equation}
where $A$ is the area of the device, and $q''_\text{solar} = Q_\text{solar}/A$. A full derivation is provided in SI Section~6.

\subsection*{Data availability}
The data supporting the conclusions of this study are included within the paper text, figures, and Supplementary Information. Any further data such as images and raw data logs are available from the corresponding author upon request.

\subsection*{Code availability}
The analysis was conducted by writing Mathematica codes that incorporate the equations and methods detailed in the paper and Supplementary Information. These codes were used to reach conclusions in the paper and are available from the corresponding author upon request.

\section*{Acknowledgements}
It is a pleasure to acknowledge Mario Mata, Isaac Berk, Bianca Navarro, Genaro Marcial-Lorza, and Vesper Evereux for assisting and designing prototype parts, and providing editing comments to the manuscript. We also acknowledge Haojie Cui and Amir Kashani for assisting and photographing in the outdoor tests as well as Donghun Pak for early-concept testing. H.J.C. acknowledges funding from the University of Nevada, Las Vegas through start-up funds, the Faculty Opportunity Award, Technology Commercialization Award, the Top Tier Doctoral Graduate Research Assistantship program, and the AANAPISI-LSAMP-McNair Summer Research Institute. H.J.C. also acknowledges support from the National Science Foundation CAREER award under Grant No. 2239416. S.Rao and N.O. acknowledge the funding provided by the Department of Defense Combat Feeding Research and Engineering Program through the Broad Agency Announcement for Basic and Applied Research at the US Army Combat Capabilities Development Command - Soldier Center through the effort ``Metal-Organic Framework (MOF) based Atmospheric Water Harvesting'' (W911QY1910010). This document is approved for public release, PAO PR2023\_81024. N.O. gratefully acknowledges the support of the National Science Foundation Graduate Research Fellowship under Grant No. 2139322.

\section*{Contributions}
Y.G. performed all experimental work with valuable assistance from A.C., S.Ricoy, A.S., and A.L. R.P. led early prototype design. N.O. and S.Rao provided DVS test results. Y.G. and H.J.C. performed data analysis and developed codes. Y.G., A.L., and H.J.C. wrote the paper. Y.G., R.P., and H.J.C. developed the concept.

\section*{Competing interests}
The authors declare no competing interests.

\bibliography{scibib}

\bibliographystyle{Science}



\clearpage

\end{document}


\maketitle

\section{Additional notes on experimental methods}
Our hydrogel-based atmospheric water harvesting (AWH) approach utilizes a thin hydrogel film at the bottom of the solution chamber. The saturated salt solution (LiBr) in the chamber creates a lower chemical potential, which provides a driving force and captures ambient water vapor. Segregated from the water storage role, the hydrogel membrane serves as a permeable medium facilitating fast capture. The scheme of our design is shown in Fig.~\ref{SM Figure 1}.

\subsection{Saturated salt solution (liquid desiccant)}
To create the low-chemical-potential environment to drive water into the liquid phase (liquid desiccant), we used a saturated aqueous solution of lithium bromide (LiBr) salt. Greenspan provided a detailed list of relative humidities of different saturated salt solutions at varied temperatures \cite{greenspan1977humidity}. LiBr provides the lowest equilibrium relative humidities compared to other salts, which would provide the largest driving force for water capture and the largest range of humidities where capture is possible. The equilibrium relative humidity of saturated LiBr solution ranges from $7.75 \pm 0.83 \%$ to $5.53 \pm 0.31 \%$ in the temperature range of \SIrange{0}{50}{\celsius}. We prepared LiBr saturated solution by adding salt into DI water gradually in an amount greater than the solubility at room temperature. During salt addition, the solution temperature increased due to the strong exothermic dissolution associated with LiBr. Using the resulting elevated temperature allowed us to ensure that the solution was fully saturated when eventually cooled to room temperature. This is because, according to the \textit{Handbook of Chemistry and Physics Online} \cite{handbook}, the solubility of LiBr (and most salts) increases with temperature. Furthermore, we confirmed saturation, as a solid phase of salt precipitated from the solution.

\subsection{Hydrogel film}

Polyacrylamide hydrogel (PAAm) films were prepared from aqueous stock solutions of the following chemicals: acrylamide (AAm, initial monomer, Merck), N,N'-methylene(bis)acrylamide (MBA, crosslinker, Sigma-Aldrich), and Irgacure 2959 (photo initiator, Sigma-Aldrich). Some important ratios were controlled during mixing: crosslinker ratio (moles of MBA over moles of AAm, \SI{0.1}{\percent}), water ratio (moles of water over moles of AAm during preparation, $11$), and initiator ratio (moles of Irgacure 2959 over moles of MBA, $0.4$). After fully dissolving all chemicals in DI water, the mixed solution was poured into a transparent mold with a fixed thickness of \SI{0.5}{\milli \meter}, equal to the thickness of the hydrogels right after synthesis. UV irradiation (\SI{365}{\nano \meter}, \SI{100}{\watt} LED array) was applied \SI{5}{\centi \meter} above the solution for \SI{1}{\hour}. Cured samples were then removed from the mold and rinsed to remove unreacted chemicals. Clean samples were immersed in DI water for \SI{3}{\day} until reaching the equilibrium state with a thickness of approximately \SI{0.7}{\milli \meter}. 

\begin{figure}[h!]
    \centering
    \includegraphics[width=\textwidth]{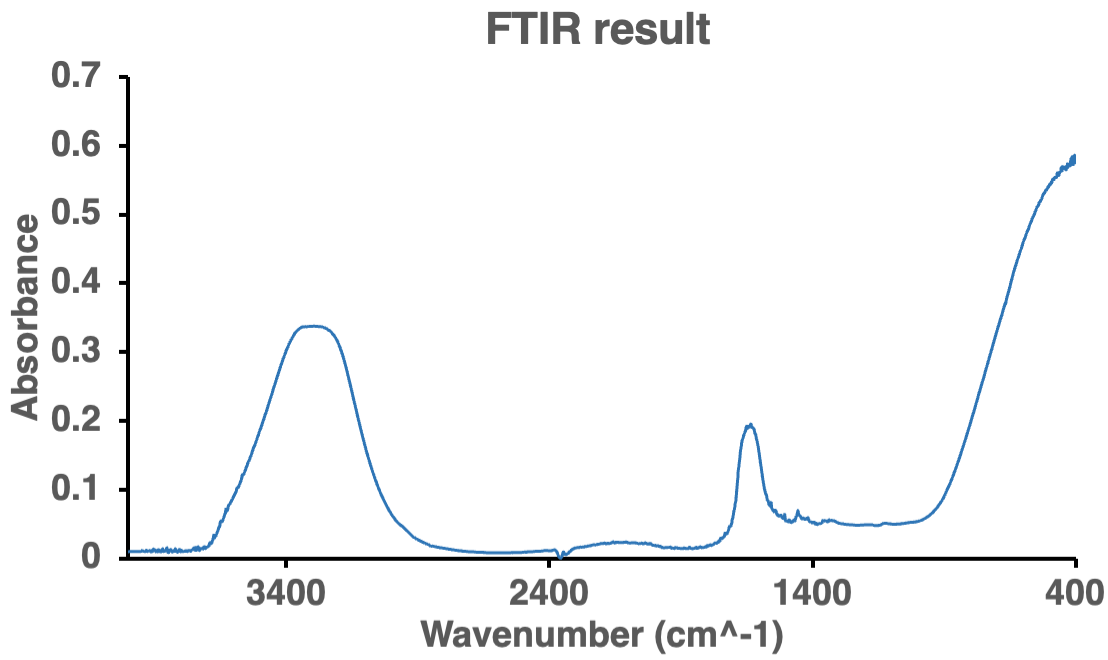}
    \caption{FTIR result of high-entanglement PAAm hydrogel sample. The crosslinker ratio of the sample is \mbox{\SI{0.1}{\percent}}}
    \label{fig.FTIR}
\end{figure}

In our previous work \protect\cite{gao2021scaling}, we varied the crosslinker ratio of pure PAAm hydrogels from \SIrange{0.5}{7}{\percent} and measured the stiffness (bulk modulus). We confirmed that adding crosslinker could increase stiffness; however, we observed an increasing brittleness. As we anticipated needing a high-strain gel with high fracture toughness, we synthesized highly entangled hydrogels with low crosslinking according to methods developed by Kim et al. \cite{kim2021fracture}. Such gels, when synthesized in a reduced water environment, have a high degree of polymer strand entanglement that greatly outnumbers crosslinks, providing high toughness, strength, and fatigue resistance. As such, we synthesized hydrogels with a crosslinker ratio = \SI{0.1}{\percent} (mol of MBA/mol of AAm), a water ratio = $11$ (mol of DI water/mol of AAm), and an initiator ratio = $0.4$ (mol of Irgacure/mol of MBA). Compared with more conventionally crosslinked gels (crosslinker ratio = \SI{0,5}{\percent}) with a bulk modulus around \SI{7}{\kilo \pascal}, our highly entangled hydrogels have a bulk modulus of \SI{28}{\kilo \pascal}. In addition, our entangled gels have a high maximum strain at failure in the range of \SIrange{160}{200}{\percent}, ensuring that our gels could withstand the high degree of stretching when constrained to a fixed area and de-swollen due to contact with the liquid desiccant. Details of the gel's response to stretching in this environment are provided in SI Section~3.

\subsection{Wind tunnel flow considerations}
For simplicity of analysis, we designed the wind tunnel to ensure laminar air flow for lab-controlled capture and storage testing. The Reynolds number, $\text{Re}$, of the air flow in the wind tunnel is
\begin{equation}
    \text{Re} = \frac{\rho_\text{air} U D_\text{h}}{\mu_\text{air}},
    \label{eq.re}
\end{equation}
where $\rho_\text{air}$ is the air density (\SI{1.205}{\kilo \gram / \meter^3}), $\mu_\text{air}$ is the dynamic viscosity of air (\SI{0.0000182}{\second \pascal}), $U$ is the velocity of air flow, and $D_\text{h}$ is the hydraulic diameter of the channel. This hydraulic diameter is
\begin{equation}
    D_\text{h} = \frac{4A}{P},
    \label{eq.dh}
\end{equation}
where $A$ is the wind tunnel cross-sectional area and $P$ is the perimeter of the tunnel ($P = 2\times (\SI{4}{\milli \meter} + \SI{38}{\milli \meter})$). By inserting the values of all parameters at our maximum attainable velocity of \mbox{\SI{0.9}{\meter\per\second}}, $\text{Re}$ of the air flow was found to be $449.7$ ($ < 2300$), which confirmed that the air flow in the wind tunnel was \textbf{laminar}. Furthermore, because we did not incorporate an entrance region before the section of the wind tunnel underneath the gel membrane, we approximated the flow conditions under the gel using flow over a flat plate with a developing boundary layer.

\subsection{Lab-controlled capture and storage tests with wind tunnel}
To test our hypothesis of convection-limited water mass transfer, we performed indoor wind-tunnel experiments underneath a prototype capture and storage device with our synthesized hydrogel membranes below a saturated LiBr solution, as shown in SI Fig.~\ref{SM Figure 1}. Tests were performed at a room temperature of $\approx \SI{23}{\celsius}$. Mass flow rates were measured under different relative humidities and wind speeds (volumetric flow rate of air). We needed to control the relative humidity and the air flow rate below the prototype (Fig.~2a) independently. Input air was supplied by our building's air supply at around \SI{5}{\percent} RH. The air flowed through a flow meter where volumetric flow rate of air input, $Q_\text{air}$, could be adjusted to a desired value using a valve---coupled with the wind tunnel, the uncertainty in velocity control was around $\pm$ \SI{0.07}{\meter\per\second}. Then, the air flow was split into two paths: in one path, air flow was kept at its dry state; in the other path, air was humidified close to \SI{99}{\percent} by bubbling through three water-filled glass media bottles. Using an Arduino microcontroller programmed with a custom PID control algorithm, we were able to adjust the ratio between dry and humid air flow to achieve a desired humidity level within $\pm$ \SI{1}{\percent} RH. The resultant, mixed air flowed beneath the hydrogel membrane via a 3D-printed wind tunnel of cross section \SI{4}{\milli \meter} by \SI{38}{\milli \meter}, with a cross-sectional area $A_\text{tunnel}$ and a flow length across the hydrogel, $W$, of \SI{38}{\milli \meter}. The average wind speed was calculated as $U = Q_\text{air}/A_\text{tunnel}$. The top wall of the wind tunnel channel was the \SI{38}{\milli \meter} by \SI{38}{\milli \meter} bottom surface of the hydrogel membrane supported by a thin metal mesh. 

A camera faced the solution chamber horizontally and was focused on the solution liquid-vapor interface to record the change in liquid level. The change in height of the solution surface was determined by image processing. Multiplying the height by the chamber cross-sectional area (\SI{38}{\milli \meter} by \SI{38}{\milli \meter}) indicated the amount of volume change, $\Delta V$ (Fig.~2b in the main text), which was then converted into captured water mass (SI Section~1E). A sensor was placed at the exit of the wind tunnel to provide feedback to the PID humidity control system. We performed $12$ independent indoor \textit{capture} tests, varying wind speed at three values (\SI{0.3}{\meter\per\second}, \SI{0.6}{\meter\per\second}, and \SI{0.9}{\meter\per\second}) and relative humidity at four values (\SI{10}{\percent}, \SI{20}{\percent}, \SI{40}{\percent}, and \SI{60}{\percent}).

A video showing the liquid volume change over time for $U=\SI{0.9}{\meter\per\second}$ at \SI{57}{\percent} RH can be viewed in the supplementary video or at \url{https://youtube.com/shorts/gomgG9pwWUQ}. The video is sped up by 750x.

\subsection{Determining the mass of captured water from measured volume change}
In our capture tests, we ensured excess LiBr salt was present to keep the solution saturated at the solubility limit even as additional water was captured and incorporated as solvent. Thus, the volume change of liquid in the chamber, $\Delta V$, measured by image processing, needed to be converted into a mass of captured water, $m_\text{capture}$. We can visualize the scenario as shown below.

\begin{figure}[h!]
    \centering
    \includegraphics[width=0.8\textwidth]{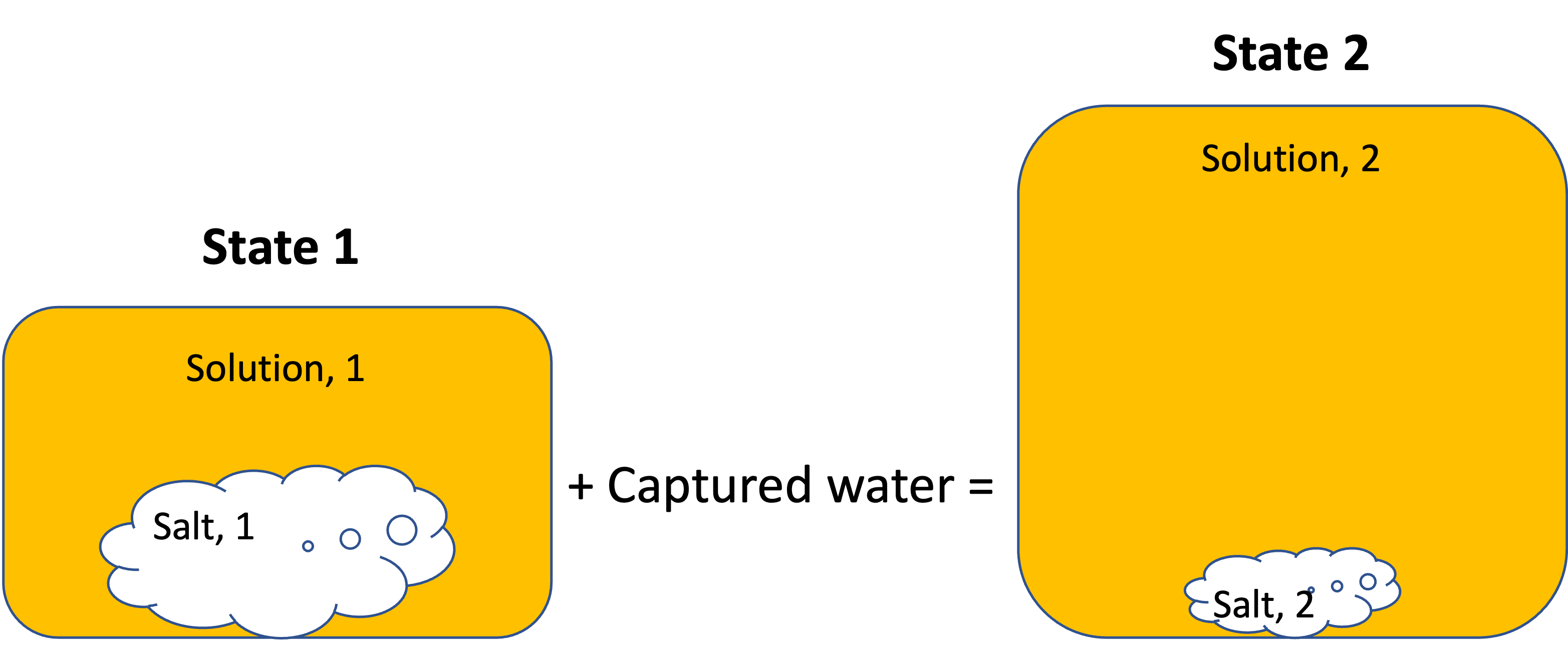}
    \caption{Before capturing water (State 1), the mass of solution is $m_\text{solution, 1}$ and the mass of LiBr solute is $m_\text{solute, 1}$; after capturing water (State 2), the mass of solution is $m_\text{solution, 2}$ and the mass of LiBr solute is $m_\text{solute, 2}$. The mass of captured water is $m_\text{capture}$.}
    \label{fig.captureprocess}
\end{figure}

The volume conservation from State 1 to State 2 can be expressed as
\begin{equation}
    \frac{m_\text{salt, 1}}{\rho_\text{salt}}+\frac{m_\text{solution, 1}}{\rho_\text{solution, 1}} + \Delta V = \frac{m_\text{salt, 2}}{\rho_\text{salt}}+\frac{m_\text{solution, 2}}{\rho_\text{solution}}{,}
    \label{eq.volumeconservation}
\end{equation}
where $\rho_\text{salt}$ is the density of the LiBr salt crystal (\SI{3.464}{\gram\per\centi\meter^3}), $\rho_\text{solution}$ is the density of the saturated LiBr solution, which depends on the temperature-dependent solubility, and $\Delta V$ is the detected volume change in the chamber. The solubility, as a mass fraction, $w$, can be calculated as a function of temperature by interpolating values from the \textit{Handbook of Chemistry and Physics Online} \cite{handbook}. Knowing the temperature and the temperature-dependent mass fraction, we can use J. M. Wimby's empirical polynomial fit \cite{wimby1994viscosity} to determine the density of LiBr solution:
\begin{align*}
    \rho_\text{solution} \left[\si{\kilogram\per\meter^3}\right] = d_\text{1}+d_\text{2}T+d_\text{3}T^2+d_\text{4}(100w)+d_\text{5}(100w)^2+d_\text{6}(100w)T\\
    +d_\text{7}(100w)^2T+d_\text{8}T^2(100w)+d_\text{9}(100w)^2+d_\text{10}(100w)^4 {.}
\end{align*}
Here, $T$ is in Celsius. The values for $d_\text{1}$ to $d_\text{10}$ are in the table below.
\begin{center}
\begin{tabular}{ |c|c| } 
 \hline
  $d_\text{1}$ & $1002.0$ \\ 
  $d_\text{2}$ & $-8.7932*10^{-2}$\\ 
  $d_\text{3}$ & $-3.79848*10^{-3}$\\ 
  $d_\text{4}$ & $8.5425$\\
  $d_\text{5}$ & $-2.9368*10^{-2}$\\
  $d_\text{6}$ & $-5.7606*10^{-3}$\\
  $d_\text{7}$ & $-8.2838*10^{-5}$\\
  $d_\text{8}$ & $7.3685*10^{-5}$\\
  $d_\text{9}$ & $1.4834*10^{-3}$\\
  $d_\text{10}$ & $4.2006*10^{-7}$\\

 \hline
\end{tabular}
\end{center}

Since the total amount of LiBr in both the solution and solid crystal is constant (no salt evaporation), the following mass conservation of salt and solute holds:
\begin{equation}
    m_\text{solute, 1} + m_\text{salt, 1} = m_\text{solute, 2} + m_\text{salt, 2} {.}
\end{equation}
Rearranging this salt mass conservation, we can define $\Delta m_\text{salt}$ as the mass change of undissolved, solid LiBr or the mass change of solute:
\begin{equation}
    \Delta m_\text{salt} = m_\text{salt, 1} - m_\text{salt, 2} = m_\text{solute, 2} - m_\text{solute, 1} \text{.}
    \label{eq.changesaltmass}
\end{equation}
We can relate the mass of solute to the mass of solution using the mass fraction, $w$, where $m_\text{solute}=w m_\text{solution}$; thus, the change in mass of solution is
\begin{equation}
    m_\text{solution,2}-m_\text{solution,1} = \frac{\overbrace{m_\text{solute,2}-m_\text{solute,1}}^{\Delta m_\text{salt}}}{w} \text{.}
    \label{eq.changesolutionmass}
\end{equation}
Combining Eqs.~\ref{eq.changesaltmass} and \ref{eq.changesolutionmass} into volume conservation, Eq.~\ref{eq.volumeconservation}, we arrive at
\begin{equation}
    \frac{\Delta m_\text{salt}}{\rho_\text{salt}} + \Delta V  = \frac{\Delta m_\text{salt}}{w\rho_\text{solution}}
    \label{eq.volumeconservation2} {.}
\end{equation}
Solving for $\Delta m_\text{salt}$,
\begin{equation}
    \Delta m_\text{salt} = \Delta V \frac{w\rho_\text{salt}\rho_\text{solution}}{\rho_\text{salt}-w\rho_\text{solution}} {.}
    \label{eq.changesalttovolume}
\end{equation}
We also know from a conservation of mass of water that
\begin{equation}
    m_\text{capture} = m_\text{solvent,2} - m_\text{solvent,1}{.}
    \label{eq.waterconservation1}
\end{equation}
By the definition of mass fraction, $w \equiv m_\text{solute}/(m_\text{solute}+m_\text{solvent})$,
\begin{equation}
    m_\text{solvent} = m_\text{solute} \left(\frac{1}{w}-1\right) \text{.}
\end{equation}
Thus, we can express the conservation of water mass, Eq.~\ref{eq.waterconservation1}, as
\begin{equation}
    m_\text{capture} = \underbrace{\left(m_\text{solute,2} - m_\text{solute,1}\right)}_{\Delta m_\text{salt}} \left(\frac{1}{w}-1\right){,}
    \label{eq.waterconservation2}
\end{equation}
where we can use Eq.~\ref{eq.changesaltmass} to express the difference in solute mass as $\Delta m_\text{salt}$. Incorporating Eq.~\ref{eq.changesalttovolume} to express $\Delta m_\text{salt}$ in terms of volume change, $\Delta V$, we arrive at our final equation relating captured water mass to observed volume change (Eq.~10 in the main text):
\begin{equation}
    m_\text{capture} = \Delta V \frac{\left(1-w\right) \rho_\text{salt}\rho_\text{solution}}{\rho_\text{salt}-w\rho_\text{solution}}{.}
    \label{eq.vtom}
\end{equation}

\section{Derivation of the convection resistance in the vapor phase}

Water vapor flows in the direction of high to low vapor pressure, which is proportional to vapor density (mass concentration), $\rho_\text{vap}$; thus, water vapor density can be used to represent vapor pressure as a driving force. The relative humidity, $\text{RH}$, is the vapor pressure, $P$, over the temperature-dependent saturation pressure, $P_\text{sat}(T)$. Combining this definition with the ideal gas law, we find that the vapor density difference between ambient and the gel-air interface, $\rho_\text{vap,amb} - \rho_\text{vap,surf}$ is proportional to a difference in ambient and relative humidity between ambient and interface, $\text{RH}_\text{amb} - \text{RH}_\text{surf}$:
\begin{equation}
    \rho_\text{vap,amb} - \rho_\text{vap,surf} = \frac{P_\text{sat}(T) M_\text{\ce{H2O}}}{RT} \left(\text{RH}_\text{amb} - \text{RH}_\text{surf}\right){,}
    \label{eq.rhdrop}
\end{equation}
where $M_\text{\ce{H2O}}$ is the molar mass of water and $R$ is the ideal gas constant.

Using a convective heat transfer analogy to describe vapor mass transfer being proportional to a difference in densities, the captured water vapor mass flux, $j$ (\si{\kilogram\per\meter\squared\per\day}), is:
\begin{equation}
    j = \frac{\dot{m}}{A} = h  (\rho_\text{vap,amb} - \rho_\text{vap,surf}) = \underbrace{h \frac{P_\text{sat}(T) M_\text{\ce{H2O}}}{RT}}_{1/(R_\text{vap}A)} \left(\text{RH}_\text{amb} - \text{RH}_\text{surf}\right){,}
    \label{eq.vspeed}
\end{equation}
where $h$ is the water vapor mass transfer coefficient with SI units of \si{\meter\per\second}. This mass transfer coefficient, $h$, can be described physically as the water vapor diffusion coefficient over the mass transfer boundary layer: $h \sim D_\text{v} / \delta$. The diffusion coefficient is $D_\text{w,a} = D_\text{w,a,0}P^{-1}T^{3/2}$, where $D_\text{w,a,0} = \SI{5.12e-4}{\meter\squared\per\second\pascal\kelvin\tothe{-3/2}}$ \cite{incropera1996fundamentals}. The thickness of the mass transfer boundary layer is related to the velocity boundary layer by the Schmidt number: $\delta/\delta_\text{v} \sim \text{Sc}^{-1/3}$ where the Schmidt number is defined as 
\begin{equation}
    \text{Sc} = \frac{\mu_\text{air} / \rho_\text{air}}{D_\text{w,a}}\text{.}
    \label{eq.sc}
\end{equation}
Since the dynamic viscosity of air is similar in magnitude to the water-vapor diffusion coefficient, $\text{Sc}\sim 1$, and the mass transfer boundary layer thickness is similar to the velocity boundary layer thickness: $\delta \sim \delta_\text{v}$. Thus, to maximize the value of captured water vapor mass flux, $j$, $\delta$ can be reduced by increasing airflow velocity (wind speed) as high as possible. The mass transfer coefficient, can be described nondimensionally by the Sherwood number, $\text{Sh}$, (analogous to the Nusselt number in heat transfer) where
\begin{equation}
    h = \text{Sh} \frac{D_\text{w,a}}{W} \text{.}
    \label{eq.h}
\end{equation}
The Blausius solution for flow over a flat plate is \cite{subramanian2014convective}
\begin{equation}
    \text{Sh} = 0.664 \text{Re}^{1/2} \text{Sc}^{1/3}{,}
    \label{eq.sh}
\end{equation}
where $W$ is the flow length (\SI{38}{\milli\meter}) of the gel. Here, we use the flat-plate Reynolds number where
\begin{equation}
    \text{Re} = \frac{\rho_\text{air} U \text{W}}{\mu_\text{air}} \text{.}
    \label{eq.re flat}
\end{equation}
Incorporating Eq.~\ref{eq.sc}--\ref{eq.re flat} into Eq.~\ref{eq.vspeed} results in an equation for vapor mass flux:
\begin{equation}
    j =  \underbrace{\frac{0.664 D_\text{w,a}^{2/3}M_\text{\ce{H2O}} P_\text{sat}\mu_\text{air}^{-1/6}\rho_\text{air}^{1/6}}{W^{1/2} R T} \sqrt{U}}_{1/(R_\text{vap}A)}(\text{RH}_\text{amb} - \text{RH}_\text{surf}) \text{.}
    \label{eq.harvestspeed}
\end{equation}

Assuming the mass-transfer resistance of the gel is small compared to that of the vapor (justified in main text), we can set $\text{RH}_\text{surf} = \text{RH}_\text{sol}$ in Eq.~\ref{eq.harvestspeed}. This allows us to calculate the water capture rate of our AWH setup at different ambient relative humidities and wind speeds. At $23^\circ C$, 
\begin{equation}
    j =  \left(\SI{3.88e-4}{\kilogram\meter\tothe{-5/2}\second\tothe{-1/2}}\right) \sqrt{U}(\text{RH}_\text{amb} - \text{RH}_\text{sol}) .
    \label{eq.harvestspeed2}
\end{equation}
From Eq.~\ref{eq.harvestspeed2} it is apparent that the water capture rate is linearly dependent on the ambient RH and the square root of wind speed. Fig.~\ref{fig.model} shows the modeled capture rate of our setup at varied wind speeds and humidities.

\section{Determination of membrane thickness, $L$}
When the hydrogel membrane is in contact with the liquid desiccant, it is in a de-swollen state due to it being equilibrated in a low-relative-humidity (low chemical potential or high osmotic pressure) environment. It is also in a stretched state since the de-swollen gel has been constrained to a fixed area equal to that of its swollen, wet (\SI{100}{\percent} RH) state. Here we calculate the thickness of the hydrogel membrane when at this de-swollen, stretched state (Fig.~\ref{fig.thickness}, bottom). We start by considering a gel at its unstretched, wet state (Fig.~\ref{fig.thickness}, top). At this state, the thickness and the width of the gel are $L_\text{wet}$ and $W$, respectively. Imagine that this gel de-swells, unconstrained (unstretched) (Fig.~\ref{fig.thickness}, middle) to a state defined by the filling fraction,
\begin{equation}
s \equiv \frac{V}{V_\text{wet}}{,}
\end{equation}
where $V$ is the volume of the de-swollen gel and $V_\text{wet}$ is the volume of the gel at the wet state. Both the new thickness, $L_s$, and width, $W_s$, at this de-swollen state are smaller than the wet-state values.

Now, consider that we take this de-swollen unstetched gel and we stretch it to its original wet-state area while keeping it de-swollen (Fig.~\ref{fig.thickness}, bottom). The stretched gel would strain. We can define the transverse infinitesimal strains (parallel to the width dimension) as
\begin{equation}
    d\varepsilon_\text{x} = d\varepsilon_\text{y} = d\varepsilon_\text{trans} = \frac{dW_s}{W_s}
    \label{eq.trans}
\end{equation}
and define the longitudinal infinitesimal strains (parallel to thickness dimension) as
\begin{equation}
    d\varepsilon_\text{z} = d\varepsilon_\text{lon} = \frac{dL_s}{L_s}.
    \label{eq.long}
\end{equation}
Poisson's ratio can be defined as 
\begin{equation}
    \nu = -\frac{\varepsilon_\text{trans}}{\varepsilon_\text{lon}}.
    \label{eq.pois}
\end{equation}
Eq.~\ref{eq.pois} can be rearranged to be 
\begin{equation}
    -\nu \underbrace{\varepsilon_\text{lon}}_{\int_{\text{unstretched}}^{\text{stretched}}d\varepsilon_\text{lon}} = \underbrace{\varepsilon_\text{trans}}_{\int_{\text{unstretched}}^{\text{stretched}}d\varepsilon_\text{trans}}{,}
    \label{eq.arrpois}
\end{equation}
where we recognize that $\varepsilon_\text{lon}$ and $\varepsilon_\text{trans}$ are integrations of the infinitesimal strains from the unstretched to stretched states of the de-swollen gel.
Plugging Eq.~\ref{eq.trans} and Eq.~\ref{eq.long} into Eq.~\ref{eq.arrpois} and integrating, 
\begin{align*}
    -\nu \int_{L_s}^{L}\frac{dL'}{L'} &= \int_{W_s}^{W}\frac{dW'}{W'}\\
    -\nu \ln\left(\frac{L}{L_s}\right) &= \ln\left(\frac{W}{W_s}\right)\\
    \ln\left(\frac{L}{L_s}\right)^{-\nu}&= \ln\left(\frac{W}{W_s}\right)\\
    \left(\frac{L}{L_s}\right)^{-\nu} &= \frac{W}{W_s}\\
    \frac{L}{L_s} &= \left(\frac{W}{W_s}\right)^{-\frac{1}{\nu}} \text{.}
\end{align*}
Solving for the thickness, we arrive at an equation of the de-swollen stretched thickness as a function of de-swollen unstretched dimensions and stretched width:
\begin{equation}
    L = L_\text{s}\left(\frac{W}{W_\text{s}}\right)^{-\frac{1}{\nu}}\text{.}
    \label{eq.thickrelation}
\end{equation}
We can relate the width of an unstretched de-swollen gel to its wet-state value by the filling fraction as
\begin{equation}
    \underbrace{W_\text{s}}_{\text{width of de-swollen gel unstretched}} = \underbrace{W}_{\substack{\text{width of stretched gel} \\ \text{(same as width of swollen unstretched gel)}}} s^{1/3} \text{.}
    \label{eq.widthswelling}
\end{equation}
We can also relate the thickness of the unstretched de-swollen gel to its wet-state value by the filling fraction as
\begin{equation}
    \underbrace{L_\text{s}}_{\text{thickness of de-swollen gel unstretched}} = \underbrace{L_\text{wet}}_{\text{thickness of stretched gel}} s^{1/3}.
    \label{eq.thickswelling}
\end{equation}
Plugging Eq.~\ref{eq.widthswelling} and Eq.~\ref{eq.thickswelling} into Eq.~\ref{eq.thickrelation},
\begin{align*}
    L &= L_\text{wet} s^{1/3} \left(\frac{W}{W s^{1/3}}\right)^{-\frac{1}{\nu}}\\
    L &= L_\text{wet} s^{1/3} s^{\frac{-1}{3} \frac{-1}{\nu}}{,}
\end{align*}
which leads to an expression for the thickness of a stretched de-swollen gel constrained to its original wet-state area:
\begin{equation}
    L = L_\text{wet} s^{\frac{1}{3} + \frac{1}{3\nu}} \text{.}
    \label{eq.thickstretch}
\end{equation}
The Poisson's ratio for all hydrogels, $\nu$, was assumed to be $1/3$ as measured previously by others for crosslinked hydrogels \cite{wyss2010capillary,Geissler1981,Andrei1998}; therefore, applying the value in Eq.~\ref{eq.thickstretch},
\begin{equation}
    L = L_\text{wet} s^{\frac{4}{3}} \text{.}
    \label{eq.thickcal}
\end{equation}
In our experiments, all hydrogel membranes for AWH tests were immersed in DI water for seven days to reach their fully wet states before use. The thickness of all membranes at their fully wet states, $L_\text{wet}$, was around \SI{0.7}{\milli\meter}. From our previous work, the filling fraction, $s$, of the hydrogel is around \SI{10}{\percent} when subjected to low humidities below \SI{50}{\percent}. As such, the thicknesses of our hydrogel membranes were  approximately \SI{0.03}{\milli\meter} when in contact with the liquid desiccant and constrained to their wet-state area.

\section{Justifying the assumption of uniform properties within the gel}
Water flows through the membrane due to a pressure gradient, which can be expressed as a gradient in filling fraction by way of the modulus of the membrane (Eq.~6 of main text). In our expression for $R_\text{gel}$ (Eq.~9 of main text) we assume that the filling fraction, $s$, is uniform within the membrane. This is an assumption we can justify by a dimensionless argument. If water flow is being applied through the membrane at a given volumetric flowrate, $Q_\text{water}$, and divided by the cross-sectional area, $A$, we have an effective velocity, $U_\text{gel}$. The characteristic timescale for this water to flow through the membrane thickness, $L$, is $t_\text{flow} = L/U_\text{gel}$. While water flows through the gel, due to the required driving forces, the gel will respond in its filling fraction gradient according to poroelastic diffusion. Thus, the timescale for the poroelastic response is $t_\text{pe}=L^2/D_\text{pe}$, where $D_\text{pe} = K \kappa /\mu$. If we compare these two timescales, a resulting poroelastic Peclet number results:
\begin{equation}
    \frac{t_\text{pe}}{t_\text{flow}}= \underbrace{\frac{L U_\text{gel}}{D_\text{pe}}}_\text{Pe} = \frac{L U_\text{gel} \mu}{K \kappa} \text{.}
\end{equation}
Assuming an applied effective velocity of $U_\text{gel}=\SI{1.15e-7}{\meter \per \second}$ (corresponding to a capture rate of \SI{10}{\kilogram \per \meter \squared \per \day}), and inserting known values of $K$, $\kappa$, $\mu$, and $L = \SI{0.03}{\milli\meter}$, we find that the resulting Peclet number is around 0.02, which is well below unity. Even if we increased the capture rate to \SI{100}{\kilogram \per \meter \squared \per \day}, well above the maximum fluxes predicted globally (Fig.~4b), the Peclet number is 0.2. Thus, our analysis of $R_\text{gel}$ should be valid for all expected fluxes anywhere.

\section{Water sorption testing}
To measure the sorption response of the PAAm hydrogels in varying humidities, the dynamic vapor sorption (DVS) technique was utilized. The DVS Adventure from Surface Measurement Systems was the machine used for these experiments. The DVS Adventure consists of a microbalance with \SI{0.01}{\micro\gram} resolution and a testing chamber with temperature and water vapor pressure control. The desired vapor pressure or relative humidity (RH) is achieved by bubbling a dry nitrogen stream through deionized (DI) water and controlled mixing with dry nitrogen using a mass flow controller. A small sample, \SI{3}{\milli\meter} by \SI{5}{\milli\meter}, was cut from the fabricated \SI{0.7}{\milli\meter} hydrogel film to fit in the measuring pan of the DVS scale. The sample was cut using a clean razor blade and handled exclusively with tweezers to prevent contamination. The hydrogel was stored in a vial of DI water to maintain its hydrated state before extraction for this experiment. Since the material began in a hydrated state a DVS methodology starting the sample chamber at \SI{98}{\percent} RH at \SI{25}{\celsius}, the closest conditions to saturation that is possible in this instrument. The chamber was maintained at \SI{25}{\celsius} for the entirety of the experiment. Starting from this condition, the sample was exposed to progressively lower RH conditions in \SI{10}{\percent} decrements with smaller changes when near the saturation point and allowed to equilibrate at each condition. Exponential fits to each step-change in relative humidity were applied to calculate the steady-state masses. From these steady-state mass values, we can construct a desorption mass isotherm. Dividing the mass isotherm by the initial wet-state mass, we obtain the mass fraction isotherm. The mass fraction isotherm is fit to a GAB isotherm model where
\begin{equation}
    \text{wf} = \frac{\frac{q_\text{m} c_\text{g} k_\text{g} \text{RH}}{(1-k_\text{g} \text{RH})(1-k_\text{g} \text{RH} + c_\text{g} k_\text{g} \text{RH})} + m_\text{poly}}{m_\text{wet}} \text{,}
\end{equation}
and $q_\text{m}$, $m_\text{poly}$ and $k_\text{g}$ are fitting parameters. $c_\text{g}$ is set to 10 based on H. Mittal's work \cite{mittal2020adsorption}. In accordance with other work \cite{icsik2004swelling,jayaramudu2019swelling,skelton2013biomimetic}, we assume similar densities between polymer and water and a relatively larger amount of water compared to polymer such that $s$ can be approximated as the weight fraction, $s \simeq wf$. By determing the value of $s$ at the solution humidity, we are able to calculate $R_\text{gel}$.

\section{Derivation of the thermodynamic limit of distillation}
The thermodynamic limit of distilling a saturated solution (Eq.~12 of the main text) can be derived by assuming a thermodynamically reversible black box with an inflow of saturated salt solution, $\dot{m}_\text{sol}$, and outflows of pure salt, $\dot{m}_\text{salt}$, and pure water, $\dot{m}_\text{water}$. Heat, $Q_\text{h}$, flows in at a hot temperature, $T_\text{h}$. Another heat, $Q_\text{c}$, flows out at a cold temperature, $T_\text{c}$. The first law around this box is 
\begin{equation}
    0 = Q_\text{h}-Q_\text{c} + \dot{m}_\text{solution} h_\text{solution} - \dot{m}_\text{salt} h_\text{salt} - \dot{m}_\text{water} h_\text{water}{,}
    \label{eq.firstlaw}
\end{equation}
where $h$ is the specific enthalpy. The second law is
\begin{equation}
    0 = \frac{Q_\text{h}}{T_\text{h}}-\frac{Q_\text{c}}{T_\text{c}} + \dot{m}_\text{solution} s_\text{solution} - \dot{m}_\text{salt} s_\text{salt} - \dot{m}_\text{water} s_\text{water} + \dot{S}_\text{gen}{,}
    \label{eq.secondlaw}
\end{equation}
where $s$ is the specific entropy. Combining the first and second law by eliminating $Q_\text{c}$ results in
\begin{equation}
    0 = Q_\text{h}\left(1 - \frac{T_\text{c}}{T_\text{h}}\right) - T_\text{c} \dot{S}_\text{gen} + \dot{m}_\text{solution} g_\text{solution} - \dot{m}_\text{salt} g_\text{salt} - \dot{m}_\text{water} g_\text{water} \text{,}
    \label{eq.combined}
\end{equation}
where $g\equiv h - T s$ is the specific Gibbs free energy. Here, we assume all mass flows are at $T_\text{c}$. The total Gibbs free energy of the solution must be
\begin{equation}
    G_\text{solution} = n_\text{solute} \mu_\text{solute} + n_\text{solvent} \mu_\text{solvent}{,}
\end{equation}
where $n$ is the moles of species and $\mu$ is the molar chemical potential. Recognizing that for a saturated salt solution, there is chemical equilibrium between the solute and pure salt,
\begin{equation}
    \mu_\text{solute} = \mu_\text{salt} \text{,}
\end{equation}
as well as between solvent and water vapor at the equilibrium relative humidity of $\text{RH}_\text{sat}$ at saturation,
\begin{equation}
    \mu_\text{solvent} = \mu_{\text{RH}_\text{sat}} \text{.}
\end{equation}
The specific Gibbs free energy is then
\begin{equation}
    g_\text{solution} = \frac{G_\text{solution}}{m_\text{solution}}= \underbrace{\frac{m_\text{solute}}{m_\text{solute}+m_\text{solvent}}}_{w} \underbrace{\frac{\mu_\text{salt}}{M_\text{salt}}}_{g_\text{salt}} + \underbrace{\frac{m_\text{solvent}}{m_\text{solute}+m_\text{solvent}}}_{1-w} \underbrace{\frac{\mu_{\text{RH}_\text{sat}}}{M_\text{water}}}_{g_{\text{RH}_\text{sat}}} \label{eq.gsol}
\end{equation}
where we recognize that $m_\text{solute}+m_\text{solvent}=m_\text{solution}$, $M$ is the molar mass, and $w\equiv m_\text{solute}/m_\text{solution}$ is the mass fraction of solute in solution. For any species $i$, the molar chemical potential is related to the specific Gibbs free energy as
\begin{equation}
    g_i = \mu_i/M_i \text{.}\label{eq.mutog}
\end{equation}
Applying this relation to the salt and the solute in equilibrium with the $\text{RH}_\text{sat}$ state,
\begin{align}
    g_\text{salt} &= \frac{\mu_\text{salt}}{M_\text{salt}}\\
    g_{\text{RH}_\text{sat}} &= \frac{\mu_{\text{RH}_\text{sat}}}{M_\text{water}} \text{.}
\end{align}
Eq.~\ref{eq.gsol} can then be simplified as 
\begin{equation}
    g_\text{solution} = w g_\text{salt} + \left(1 - w\right) g_{\text{RH}_\text{sat}}{.}
\end{equation}
Substituting the above into Eq.~\ref{eq.combined} along with the relation of $\dot{m}_\text{salt}/\dot{m}_\text{water} = w/\left(1-w\right)$ results in
\begin{equation}
    \frac{Q_\text{h}}{\dot{m}_\text{water}}\left(1-\frac{T_\text{c}}{T_\text{h}}\right) = g_\text{water} - g_{\text{RH}_\text{sat}} + T_\text{c} \dot{S}_\text{gen} \text{.}
    \label{eq.combined2}
\end{equation}
The pure water coming out is in chemical equilibrium with water vapor at \SI{100}{\percent} relative humidity, so
\begin{equation}
    g_\text{water} = \frac{\mu_{\text{RH}_\text{100\%}}}{M_\text{water}} \text{.}
\end{equation}
Assuming the following chemical potential for water vapor,
\begin{equation}
    \mu_\text{RH} = \mu_{\text{RH}_\text{100\%}} + R T \ln \left(\text{RH}\right){,}
\end{equation}
then Eq.~\ref{eq.combined2} is
\begin{equation}
    \frac{Q_\text{h}}{\dot{m}_\text{water}}\left(1-\frac{T_\text{c}}{T_\text{h}}\right) = \frac{R T_\text{c} \ln \left(\frac{1}{\text{RH}_\text{sat}}\right)}{M_\text{water}} + T_\text{c} \dot{S}_\text{gen} \text{.}
\end{equation}
For a thermodynamically reversible system, $\dot{S}_\text{gen} = 0$. Furthermore, in the ideal case, the cold temperature reservoir is the ambient temperature, $T_\text{c} = T_\text{amb}$. Thus, the thermodynamic limit of distillation is
\begin{equation}
    \frac{Q_\text{h}}{\dot{m}_\text{water}}\left(1-\frac{T_\text{amb}}{T_\text{h}}\right) = \frac{R T_\text{amb} \ln \left(\frac{1}{\text{RH}_\text{sat}}\right)}{M_\text{water}}  \text{.}
\end{equation}
Note that the quantity of $Q_\text{h}\left(1-\frac{T_\text{amb}}{T_\text{h}}\right)$ is the work from an ideal Carnot heat engine. For the purpose of estimating a reasonable maximum heat source temperature in Fig.~4c of the main text, we set $T_\text{h}$ to \SI{373}{\kelvin} corresponding to the boiling point of water. Assuming solar-powered release where $Q_\text{h} = Q_\text{solar}$, we quantify the thermodynamic limit of water mass flux, $j_\text{III}$, as 
\begin{equation}
    j_\text{III} = \frac{\dot{m}_\text{water}}{A} = q''_\text{solar}\left(1-\frac{T_\text{amb}}{T_\text{h}}\right)\frac{M_\text{water}}{R T_\text{amb} \ln \left(\frac{1}{\text{RH}_\text{sat}}\right)}{,}
\end{equation}
where $A$ is the area of the device, and $q''_\text{solar} = Q_\text{solar}/A$ is the solar irradiation. 

\section{Determination of the convection resistance in solution, $R_\text{sol}$}
The resistances associated with transport from the inner surface of the membrane to and through the desiccant solution, $R_\text{sol}$, can be explained by the analysis of free convection in an enclosed space, where we can introduce a solutal Rayleigh number, $\text{Ra}$, that quantifies the rate of solute-concentration-induced buoyancy-driven mixing (SBM) over the rate of molecular diffusion \cite{bratsun2021mechanisms}.
\begin{equation}
    \begin{aligned}
    \text{Ra} = {}& \frac{\text{rate of SBM}}{\text{rate of diffusion}} = \frac{t_\text{diffusion}}{t_\text{SBM}} = \frac{L^2/D_\text{LiBr}}{\frac{\mu}{\Delta \rho Lg}} = \frac{\Delta \rho g L^3}{\mu D_\text{LiBr}} = \frac{\frac{\partial \rho}{\partial w} \Delta w g L^3}{\mu D_\text{LiBr}} \\
    & = \frac{\frac{\partial \rho}{\partial w} |\frac{\partial w}{\partial \text{RH}}| \Delta \text{RH} g L^3}{\mu D_\text{LiBr}} = \frac{\frac{\partial \rho}{\partial w} \frac{w_\text{sat}}{1-\text{RH}_\text{sat}} (\text{RH}_\text{gel--sol} - \text{RH}_\text{sol}) g L^3}{\mu D_\text{LiBr}} {.}
    \end{aligned}
\end{equation}
Here, the driving force is the difference in equilibrium humidities between the top of the gel surface and the bulk solution, $\text{RH}_\text{gel--sol} - \text{RH}_\text{sol}$. The characteristic length scale, $L$, is typically given as the height of the liquid solution. The relation $|\frac{\partial w}{\partial \text{RH}}| = \frac{w_\text{sat}}{1-\text{RH}_\text{sat}}$ comes from the fact that the salt concentration is nearly linear with the drop in $\text{RH}$ \cite{bowler2017raoult}, which is a consequence of Raoult's law, thus, 

\begin{equation}
    \text{RH} = 1-(1-\text{RH}_\text{sat})\frac{w}{w_\text{sat}}{,}
\end{equation}
where $w$ is the weight fraction of salt, $w_\text{sat}$ is the weight fraction at the solubility limit (saturated state), and $\text{RH}_\text{sat}$ is the equilibrium relative humidity at the saturated state. This relation has the appropriate limits, where $\text{RH} = 1$ when $w = 0$, and $\text{RH} = \text{RH}_\text{sat}$ when $w=w_\text{sat}$. Solving for $w$,

\begin{equation}
    w = \frac{(1-\text{RH})w_\text{sat}}{1-\text{RH}_\text{sat}}
\end{equation}

The density derivative with weight fraction, $\frac{\partial \rho}{\partial w}$, can be calculated from the empirical relation given by Wimby \& Berntsson. The diffusion coefficient can be calculated fromt the CRC handbook \cite{haynes2016crc} as

\begin{equation}
    D_\text{LiBr} = \frac{(z_\text{+} + |z_\text{-}|)D_\text{+}D_\text{-}}{z_\text{+}D_\text{+} + |z_\text{-}|D_\text{-}} = \frac{2 D_\text{Li\textsuperscript{+}} D_\text{Br\textsuperscript{-}}}{D_\text{Li\textsuperscript{+}} + D_\text{Br\textsuperscript{-}}} {,}
\end{equation}

thus, $D_\text{LiBr}$ is calculated as $1.377*10^{-9}$ \si{\meter ^2 \per \second}. 

We define $\rho = \rho_\text{w} + \rho_\text{s}$, then the water flux within the solution can be calculated as 
\begin{equation}
    \begin{aligned}
    j_\text{sol} = {}& h_\text{sol}(\rho_\text{w, gel--sol} - \rho_\text{w, sol}) = h_\text{sol} \rho (\frac{\rho_\text{w, gel--sol}}{\rho}-\frac{\rho_\text{w,sol}}{\rho}) \\
    & = h_\text{sol}\rho ((10w_\text{gel--sol})-(1-w_\text{sol})) = h_\text{sol}\rho (w_\text{sol}-w_\text{gel--sol}) \\
    & = h_\text{sol}\rho \left(\frac{(1-\text{RH}_\text{sol})w_\text{sat}}{1-\text{RH}_\text{sat}} - \frac{(1-\text{RH}_\text{gel--sol})w_\text{sat}}{1-\text{RH}_\text{sat}}\right) \\
    & = \underbrace{\frac{h_\text{sol} \rho w_\text{sat}}{1- \text{RH}_\text{sat}}}_{1/(R_\text{sol}A)} (\text{RH}_\text{gel--sol}-\text{RH}_\text{sol}) {.}
    \end{aligned}
\end{equation}

We can nondimensionalizing the mass tranfer coefficient, $h_\text{sol}$, in terms of a Sherwood number (analogous to Nusselt number in heat transfer) as
\begin{equation}
    \text{Sh} \equiv \frac{h_\text{sol}L}{D_\text{LiBr}} {.}
\end{equation}

Unsing a mass transfer analogy to natural convection, the Sherwood number is related to the Rayleigh number according to the Globe and Dropkin correlation (appropriate for the first approximation), where $\text{Sh} = 0.069 \text{Ra}^{1/3}\text{Sc}^{0.074}$.
Thus, the mass transfer coefficient is 
\begin{equation}
    h_\text{sol} = {\overbrace{0.069{\underbrace{\left(\frac{\frac{\partial\rho}{\partial w}\frac{w_\text{sat}}{1-\text{RH}_\text{sat}}\left(\text{RH}_\text{gel--sol}-\text{RH}_\text{sol}\right)gL^3}{da}\right)}_{\text{Ra}}} ^{1/3} {\underbrace{\left( \frac{\mu / \rho}{D_\text{LiBr}}\right)}_{\text{Sc}}}^{0.074}}^{\text{Sh}}}\frac{D_\text{LiBr}}{L} {.}
\end{equation}
and thus, the total solution flux is 
\begin{equation}
    j_\text{sol} = \underbrace{{\overbrace{0.069{\underbrace{\left(\frac{\frac{\partial\rho}{\partial w}\frac{w_\text{sat}}{1-\text{RH}_\text{sat}}\left(\text{RH}_\text{gel--sol}-\text{RH}_\text{sol}\right)gL^3}{da}\right)}_{\text{Ra}}} ^{1/3} {\underbrace{\left( \frac{\mu / \rho}{D_\text{LiBr}}\right)}_{\text{Sc}}}^{0.074}}^{\text{Sh}}}\frac{D_\text{LiBr}}{L} \frac{\rho w_\text{sat}}{1-\text{RH}_\text{sat}}}_{1/(R_\text{sol}A)}\left(\text{RH}_\text{gel--sol} - \text{RH}_\text{sol}\right) {.}
\end{equation}

Therefore, we can introduce our electric circuit analogy and obtain a direct relation between the solution flux and $R_\text{sol}$ as
\begin{equation}
    \frac{j_\text{sol}}{\text{RH}_\text{gel--sol} - \text{RH}_\text{sol}}A = 1/R_\text{sol} {,}
\end{equation}

and calculate $R_\text{sol}$ as
\begin{equation}
    R_\text{sol} = \frac{\text{RH}_\text{gel--sol}-\text{RH}_\text{sol}}{j_\text{sol}A} {.}
\end{equation}
Calculating the resistance for saturated LiBr solution at \SI{23}{\celsius} where $\text{RH}_\text{gel--sol} = 0.1$ (a very modest RH difference between the gel top and the bulk liquid concentration), we can obtain a value of the solution resistance for our device being around $\text{R}_\text{sol} = 0.093*10^6 \si{\second \kilogram^{-1}}$, which is at least an order of magnitude lower than $R_\text{vap}$. Thus, the solution resistance can be neglected. To understand why this resistance is so low, we can look at the Rayleigh number, which is on the order of $10^9$ when $\text{RH}_\text{gel--sol} = 0.1$. This Rayleigh number is well above the critical Rayleigh number of 1708 \cite{incropera1996fundamentals}; thus, Rayleigh-Benard convection cells develop, causing buoyancy-driven mixing to occur within the solution. 

\begin{figure}[htb]
    \centering
    \includegraphics[width=\textwidth]{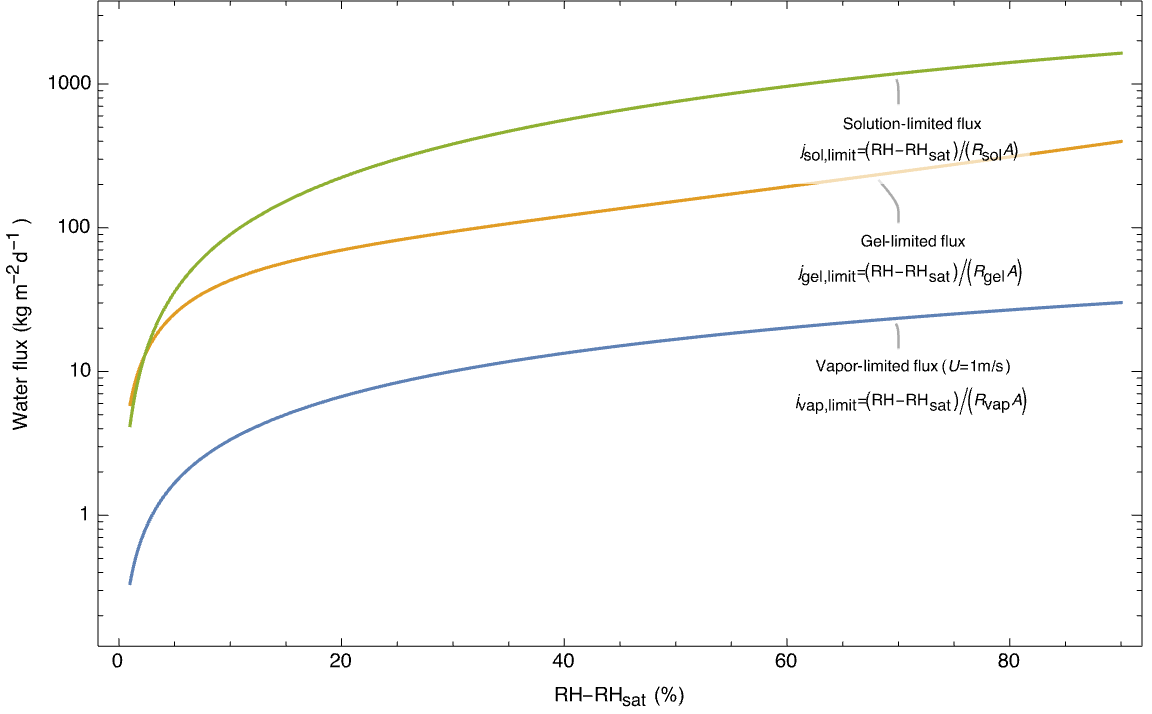}
    \caption{Comparing vapor-limited flux, gel-limited flux and solution-limited flux, we can find the convection-limited resistance in the solution,$R_\text{sol}$, is significantly lower than both $R_\text{gel}$ and $R_\text{vap}$. Thus, $R_\text{sol}$ can be neglected during the analysis of the water capture flux.}
    \label{fig.3resistance}
\end{figure}

\section{Determination of the specific latent heat of vaporization, $h_\text{fg}$}

The enthalpy of vaporization (latent heat) will also change with the addition of salt owing to the fact that the enthalpy of water in both liquid and equilibrium vapor states will change in the presence of salt \cite{sharqawy2010thermophysical}. We are interested in the change of latent heat specific to pure water mass (as opposed to total solution mass) since this will provide the quantity of energy to vaporize water per unit of produced water. The specific enthalpy of phase change per unit mass of pure water is then

\begin{equation}
    h_\text{w, fg} = h_\text{w, v} - h_\text{w, l} {,}
\end{equation}

where $h_\text{w, v}$ is the latent heat of vapor and $h_\text{w, l}$ is the latent heat of liquid. 

The specific change of solution Gibbs free energy with the addition of salt is 

\begin{equation}
    \Delta g_\text{mix} = \Delta h_\text{mix} - T \Delta s_\text{mix} {.}
\end{equation}

As the equilibrium vapor pressures \cite{kaita2001thermodynamic} at the solubility limit are known (CRC handbook), it is possible to determine the specific Gibbs free energy change for water where

\begin{equation}
    \Delta g_\text{w} = g_\text{w, v} (T, P_\text{vap}) - g_\text{w, v} (T, P_\text{sat}) {.}
\end{equation}

Note that difference in $g_\text{w}$ arises from the fact that when there is only pure water, the liquid is in equilibrium with water vapor at vapor pressure, $P_\text{sat}$; thus, the specific Gibbs free energies are equal: $g_\text{w, l, pure} = g_\text{w, v} (T, P_\text{sat})$. When salt is added, the water in the liquid phase is now in equilibrium with the vapor phase a reduced vapor pressure of $P_\text{vap}$ such that $g_\text{w, l} = g_\text{w, v} (T, P_\text{vap})$.

Since the aprtial specific Gibbs free energy change is $\Delta g_\text{w} = (1-w) \Delta g_\text{mix}$, where w is the weight fraction of salt, then
\begin{equation}
    \Delta g_\text{w} = (1-w) \Delta h_\text{mix} - T \Delta s_\text{mix} {.}
\end{equation}
Assuming ideal mixing, which is reasonable for an estimation of latent heat since the entropic contribution to $\Delta g_\text{mix}$ is quite small, 

\begin{equation}
    \Delta s_\text{mix} = \frac{-R(x_\text{w} \ln(x_\text{w} + x_\text{Li} \ln(x_\text{Li} + x_\text{Br} \ln(x_\text{Br})}{M_\text{avg}} 
\end{equation}

where $R$ is the molar gas constant, $x_\text{i}$ is the mole fraction of species $i$, and $M_\text{avg}$ is the average molar mass. 

Since $\Delta s_\text{mix}$, $T$, $w$ and $\Delta g_\text{w}$ are known or can be calculated, $\Delta h_\text{mix}$ can be determined. From here, the latent heat per unit of pure water mass is 
\begin{equation}
    h_\text{w, fg} = h_\text{w, v}(T, P_\text{vap} - (h_\text{w, l}(T) + (1-w) \Delta h_\text{mix}).
\end{equation}
The results are plotted below. The latent heat is increased compared to water by up to around \SI{20}{\percent}.

\begin{figure}[htb]
    \centering
    \includegraphics[width=\textwidth]{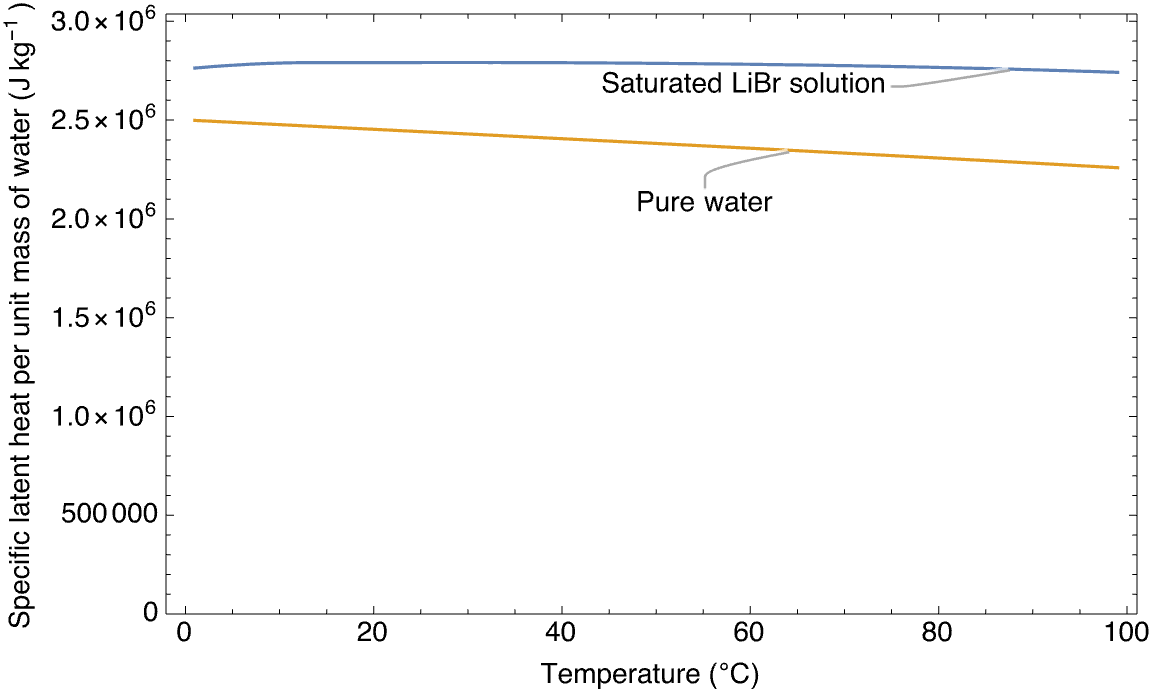}
    \caption{Specific latent heat of evaporization, $h_\text{fg}$, of LiBr solution vs. pure water. Compared to pure water, the latent heat of LiBr is increased by up to $\approx$ \SI{20}{\percent}.}
    \label{fig:my_label}
\end{figure}

\section{Calculation of water uptake for lithium bromide solution}
Uptake, defined as mass of water absorbed over mass of sorbent, is a common measure of storage performance for AWH systems. It is arguable how valuable uptake is for real performance since many systems could be transport-limited rather than storage-limited (e.g., diffusion of water in and out of a sorbent material or heat transfer limitations per unit area given solar irradiation). In any case, the uptake performance of our system can be quantified since the sorbent is a common salt---lithium bromide. Note that the mass of the gel membrane is less than one percent of the required mass of salt to store \SI{10}{\kilogram\per\meter\squared} of water; thus, we can reasonably neglect the gel mass from the uptake calculation. Using the definition of the mass fraction of salt, 
\begin{equation}
    w=\frac{m_\text{salt}}{m_\text{salt}+m_\text{water}} \text{,}
\end{equation}
and the definition of uptake, $Q_\text{uptake}$,
\begin{equation}
    Q_\text{uptake} = \frac{m_\text{water}}{m_\text{salt}} \text{,}
\end{equation}
we can eliminate the mass terms and find an expression for uptake in terms of the salt weight fraction:
\begin{equation}
    Q_\text{uptake} = \frac{1-w}{w} \text{.}
\end{equation}
Assuming a linear dependence of salt weight fraction with relative humidity \cite{bowler2017raoult}, the salt weight fraction can be expressed in terms of the salt weight fraction at the solubility limit, $w_\text{sat}$, the relative humidity at the solubility limit, $\text{RH}_\text{sat}$, and the equilibrium relative humidity, $\text{RH}$:
\begin{equation}
    w = \frac{(1-\text{RH})w_\text{sat}}{1-\text{RH}_\text{sat}} \text{.}
\end{equation}
Substituting this into the equation for uptake, we obtain an expression for the uptake in terms of the relative humidity of the salt at the solubility limit, $\text{RH}_\text{sat}$ (data available \cite{greenspan1977humidity}), the ambient relative humidity, $\text{RH}$, and the weight fraction at solubility limit, $w_\text{sat}$ (data available \cite{haynes2016crc}):
\begin{equation}
    Q_\text{uptake} = \frac{1-\text{RH}_\text{sat}}{(1-\text{RH})w_\text{sat}}-1 \text{.}
\end{equation}
Plotting the uptake as a function of ambient relative humidity for \SI{25}{\celsius}, we can see that the uptake performance of lithium bromide is comparable to even the highest-uptake gel-based sorbents in the literature \cite{guo2022scalable,graeber2023extreme}. This result is unsurprising since lithium, with its small ionic size, can form extremely tight solvation shells with high hydration energy, providing a very high affinity to water \cite{kubota2009selectivity}. 
\begin{figure}[h]
    \centering
    \includegraphics[width=0.75\textwidth]{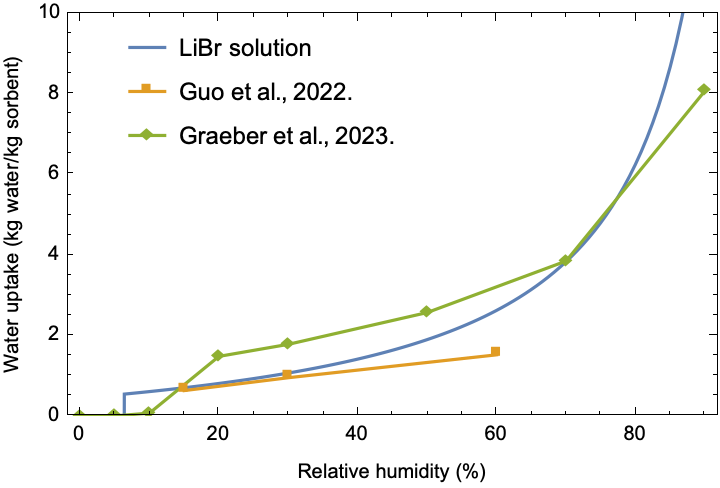}
    \caption{Water uptake of lithium bromide solution as a function of ambient humidity at \SI{25}{\celsius} exceeds the state of the art in gel-based sorbents \cite{guo2022scalable,graeber2023extreme}. Note that the discontinuous step around 6\% RH is due LiBr solution hitting the solubility limit \cite{greenspan1977humidity}.}
\end{figure}

\clearpage

\section{Figures}

\begin{figure*}[h]
    \centering
    \includegraphics[width=0.5\textwidth]{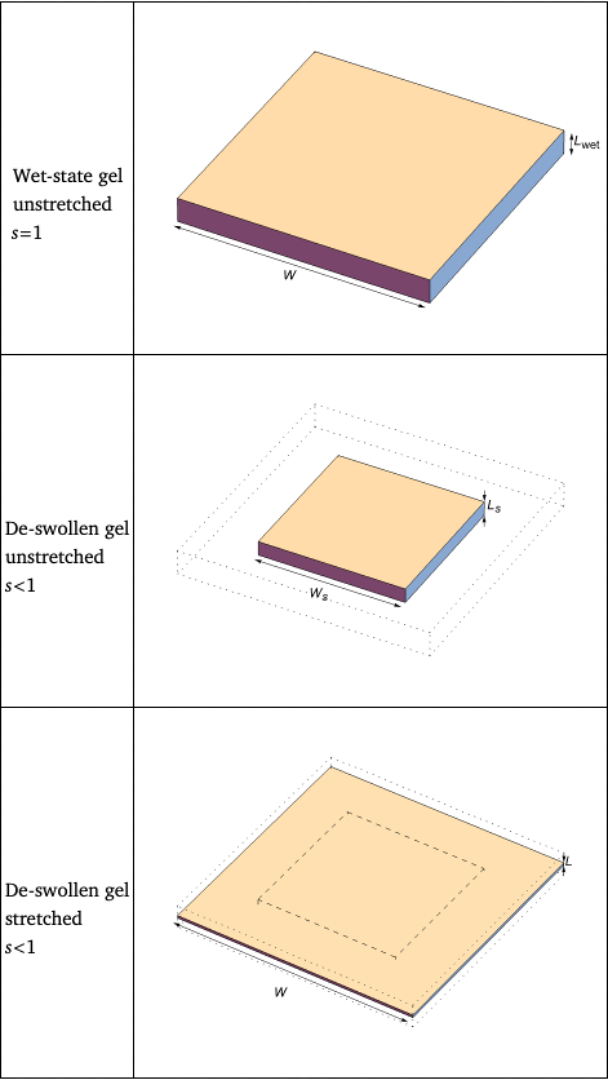}
    \caption{Illustrations of hydrogel membranes at the (top) wet unstretched, (middle) de-swollen unstretched, and (bottom) de-swollen stretched states.}
    \label{fig.thickness}
\end{figure*}

\begin{figure}[h]
    \centering
    \includegraphics[width=\textwidth]{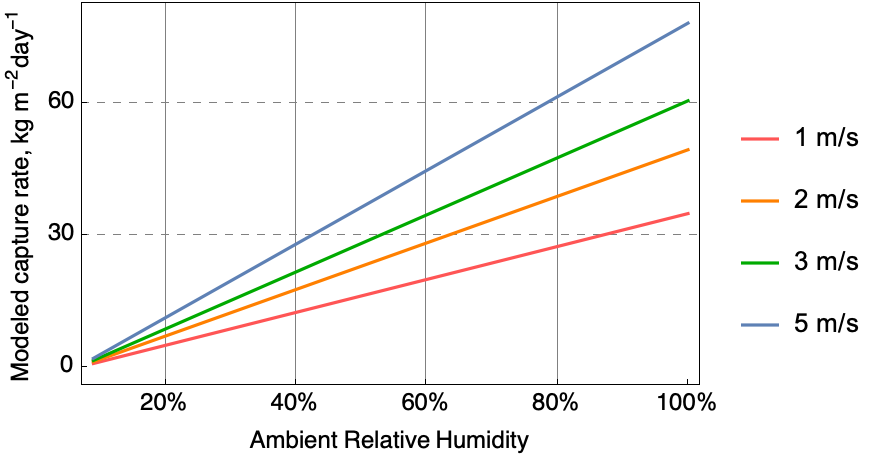}
    \caption{Expected convection-limited water capture flux at $\approx 23^\circ C$ (lab conditions) at different air velocities and humidities.}
    \label{fig.model}
\end{figure}

\begin{figure}
    \centering
    \includegraphics[width=\textwidth]{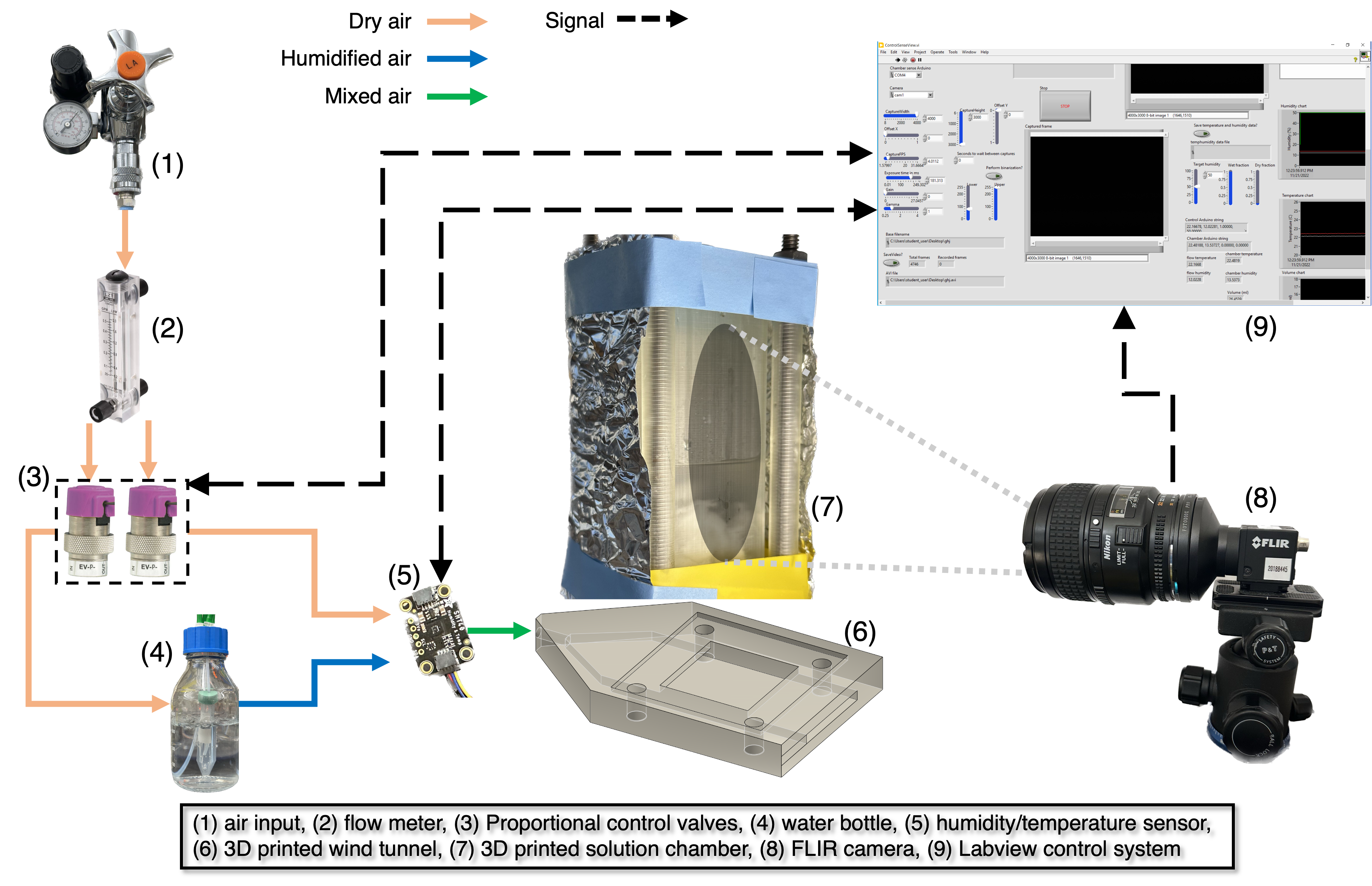}
    \caption{Details of lab-controlled water capture setup. We used a flow meter (2) to control the wind speed into the (6) wind tunnel. (3) Proportional control valves were connected to an Arduino board with a PID algorithm. The PID system adjusted the opening of both valves based on the reading of the (5) humidity sensor. The (8) camera photographed the liquid level in the (7) solution chamber every \SI{30}{s}, and the digital image was processed by (9) a computer using a combination of Mathematica and LabView in realtime.}
    \label{SM Figure 1}
\end{figure}

\begin{figure}
    \centering
    \includegraphics[width=0.8\textwidth]{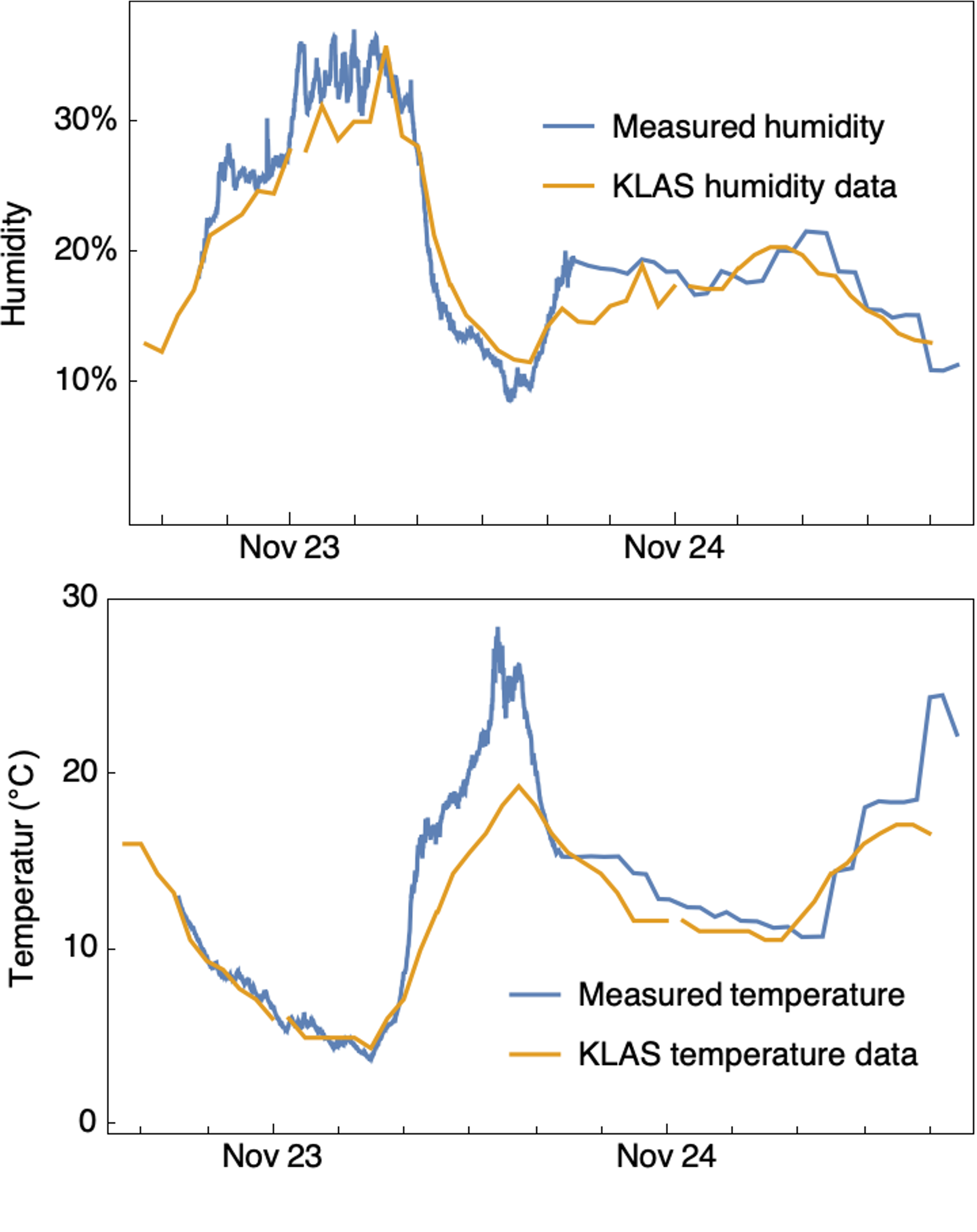}
    \caption{We compared our sensor reading on temperature and humidity (blue curves) and the weather data from the nearby KLAS airport obtained from Wolfram Research \cite{wolframweather} (orange curves), and confirmed that our measurements of the ambient environment during outdoor tests were reliable.}
    \label{SM Figure 2}
\end{figure}

\begin{figure}
    \centering
    \includegraphics[width=\textwidth]{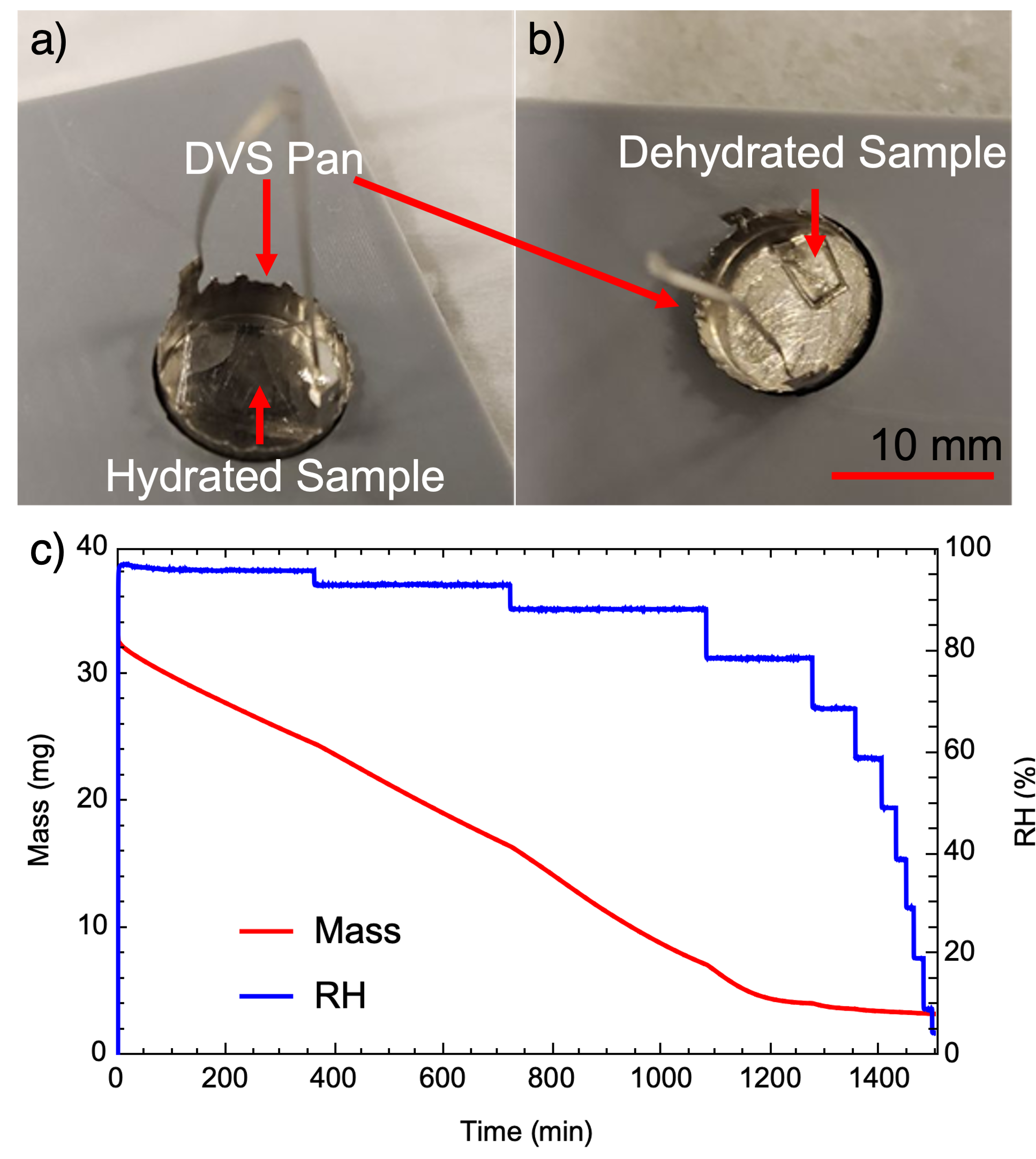}
    \caption{(a) The hydrated sample being loaded into the sample pan of the DVS before the experiment. (b) After the initial desorption, the hydrogel sample loses significant volume. (c) The RH for the sample chamber and the mass response of the hydrogel sample during the DVS experiment. The large loss of mass during the drying steps corresponds to the significant volume changes of the material.}
    \label{SM isotherm}
\end{figure}

\begin{figure}
    \centering
    \includegraphics[width=\textwidth]{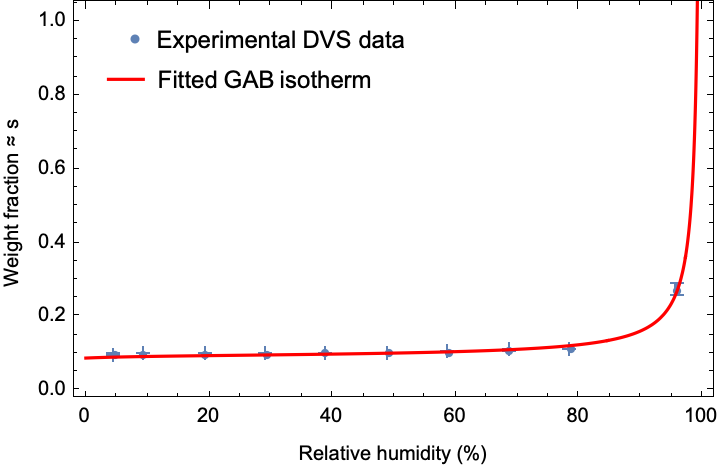}
    \caption{The GAB isotherm was fit to experimental DVS data for the weight fraction isotherm describing the degree of swelling of the gel as a function of relative humidity at \mbox{\SI{25}{\celsius}}.}
    \label{GAB fit}
\end{figure}

\begin{figure}
    \centering
    \includegraphics[width=\textwidth]{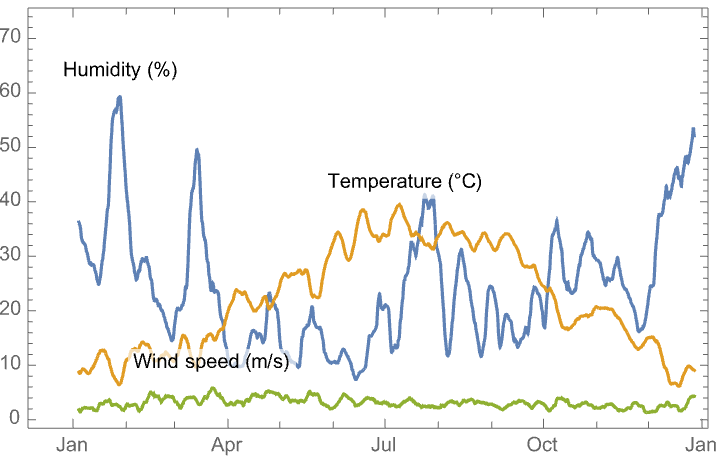}
    \caption{Weather data in 2021 from nearby KLAS airport obtained from Wolfram Research \cite{wolframweather}. Humidity is shown in blue, temperature is shown in orange and wind speed is shown in green.}
    \label{fig.KLAS rh t ws}
\end{figure}

\begin{figure}
    \centering
    \includegraphics[width=\textwidth]{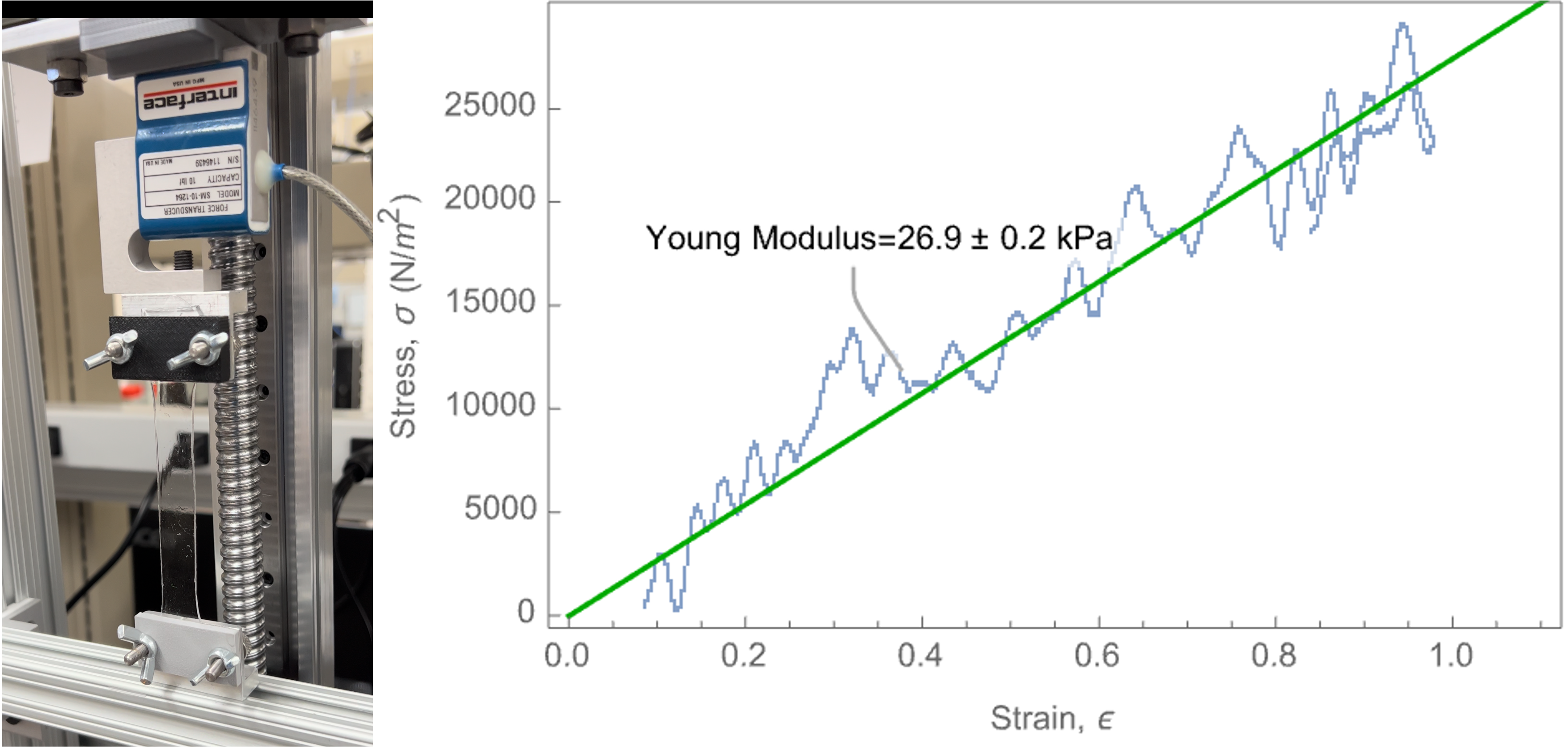}
    \caption{An example of tensile test of PAAm hydrogel used in this paper. We applied six individual tensile tests with different samples and calculated the average value and uncertainty. The sample is prepared in dogbone shape. The dimension of the stretching section is \SI{6.6}{\milli\meter} wide and \SI{1.7}{\milli\meter} thick.}
    \label{fig.tensiletestexample}
\end{figure}

\begin{figure}
    \centering
    \includegraphics[width=\textwidth]{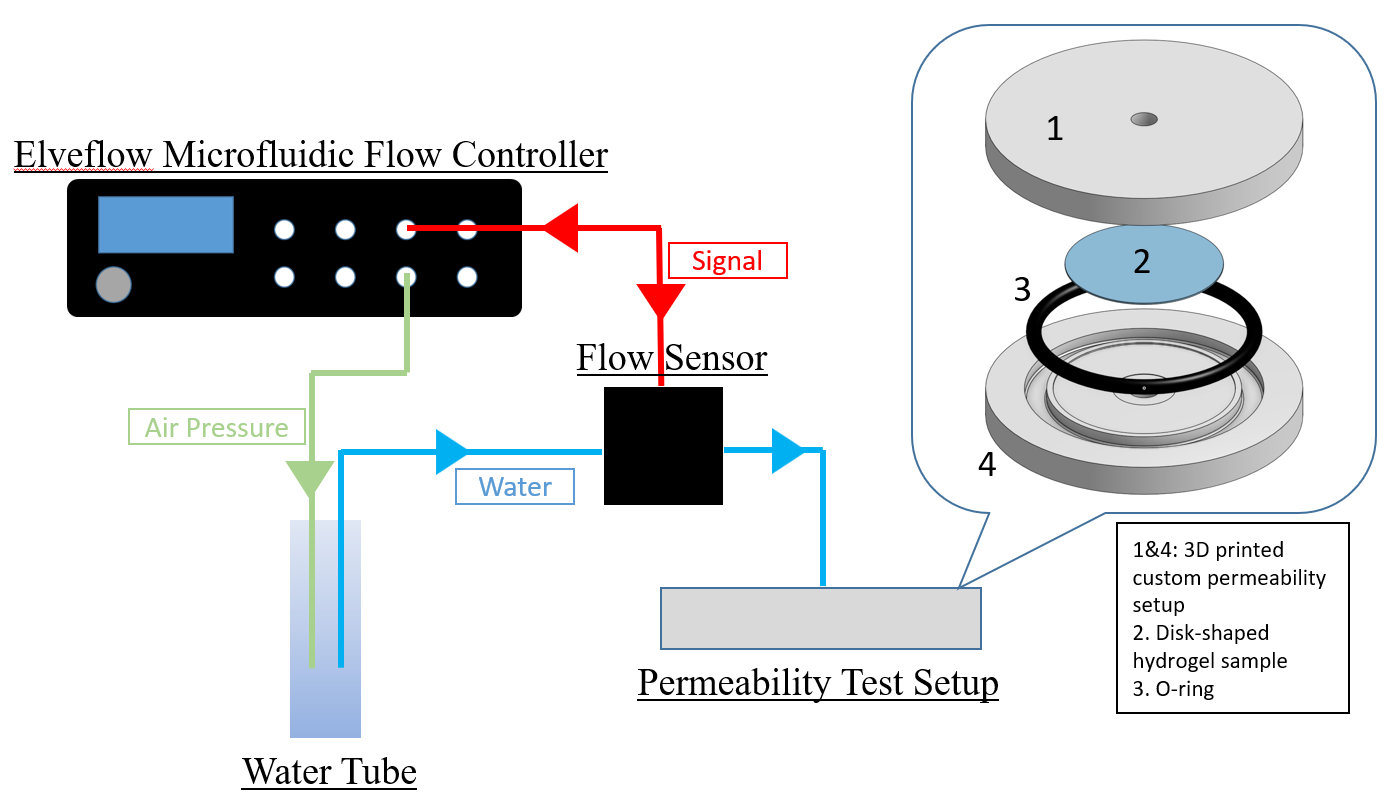}
    \caption{A scheme of the custom-built permeability tester. The custom permeability setup was designed and printed in our lab. With the setup, the gel thickness, $L$, and cross-sectional area, $A$, can be fixed at certain values. To ensure the sealing of the setup, an o-ring was applied between top and bottom parts. After the sample was inserted in the setup, two clamps were used to keep the setup tight through the testing. $\Delta P$ was controlled by the Elveflow Microfluidic Flow Controller. Water in the water tube was pushed out by the air pressure and water flowed through the flow sensor, where a reading of the volumetric flow rate, $Q$, was measured. Using Darcy's Law, we were able to determine the permeability.}
    \label{fig.permeabilitysetup}
\end{figure}

\begin{figure}
    \centering
    \includegraphics[width=\textwidth]{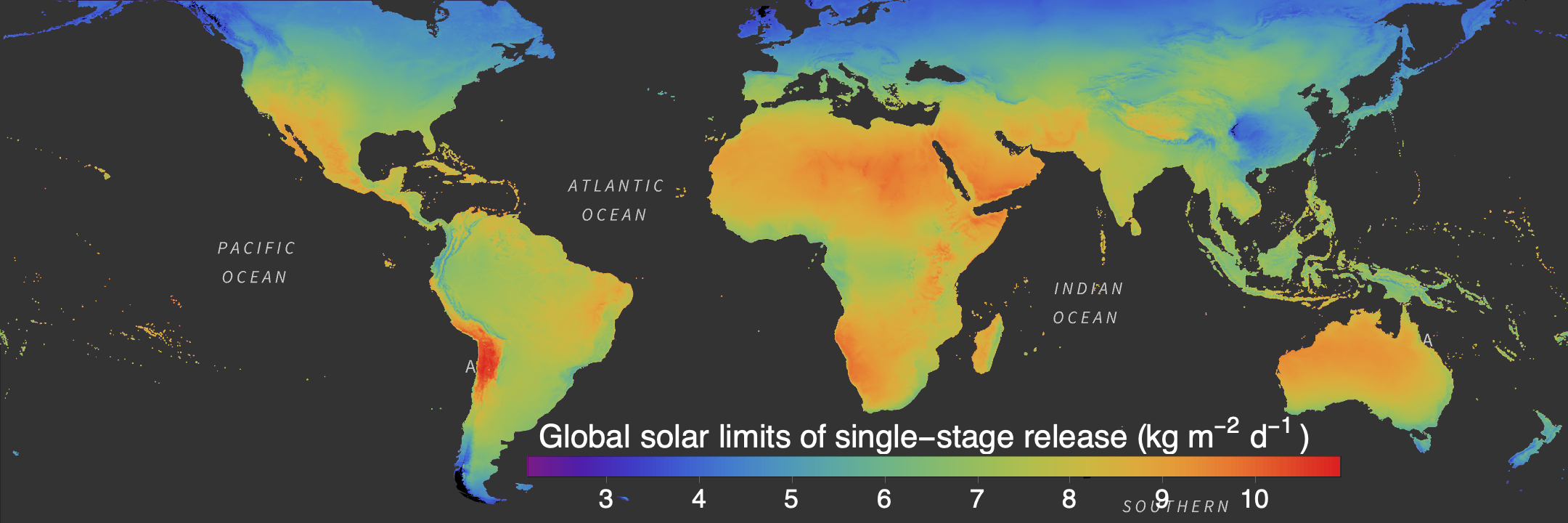}
    \caption{Global solar limits of single-stage release, $j_\text{II}$, assuming complete conversion of global horizontal irradiation \cite{gsa} into latent heat (Eq.~1 of main text).}
    \label{fig:solarlimit}
\end{figure}

\begin{figure}
    \centering
    \includegraphics[width=\textwidth]{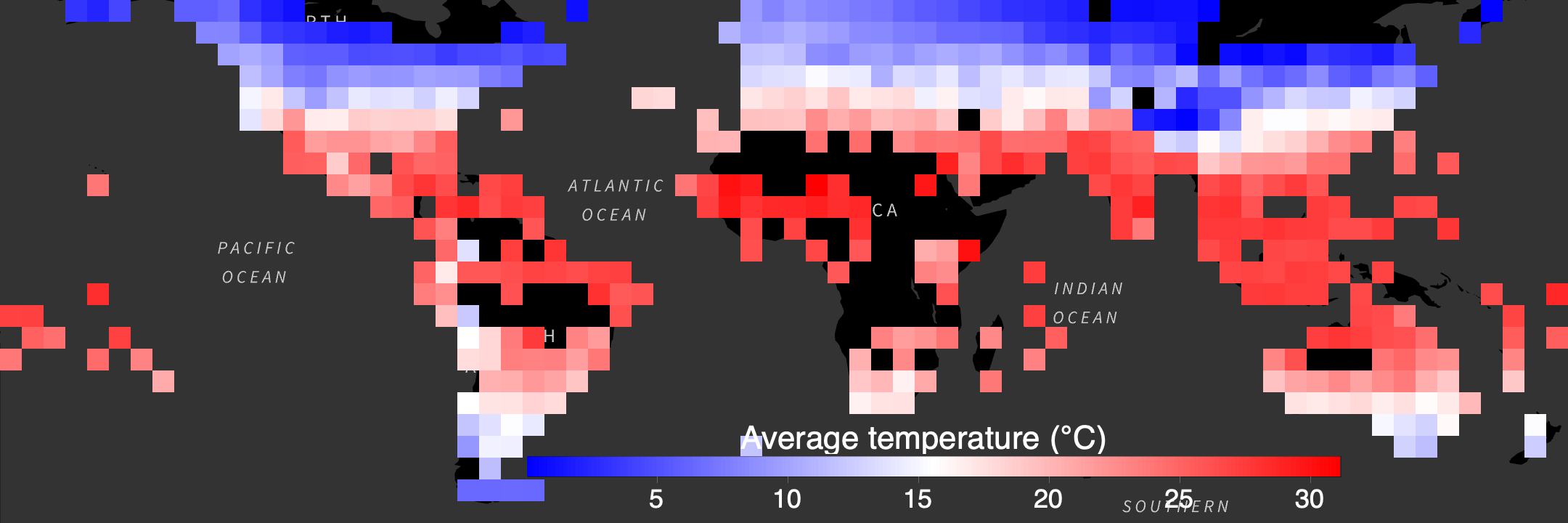}
    \caption{Global temperature data averaged over year 2021 from \cite{hadisdh}. Note that missing pixels are due to incomplete data in the original data set.}
    \label{fig:temperatures}
\end{figure}

\begin{figure}
    \centering
    \includegraphics[width=\textwidth]{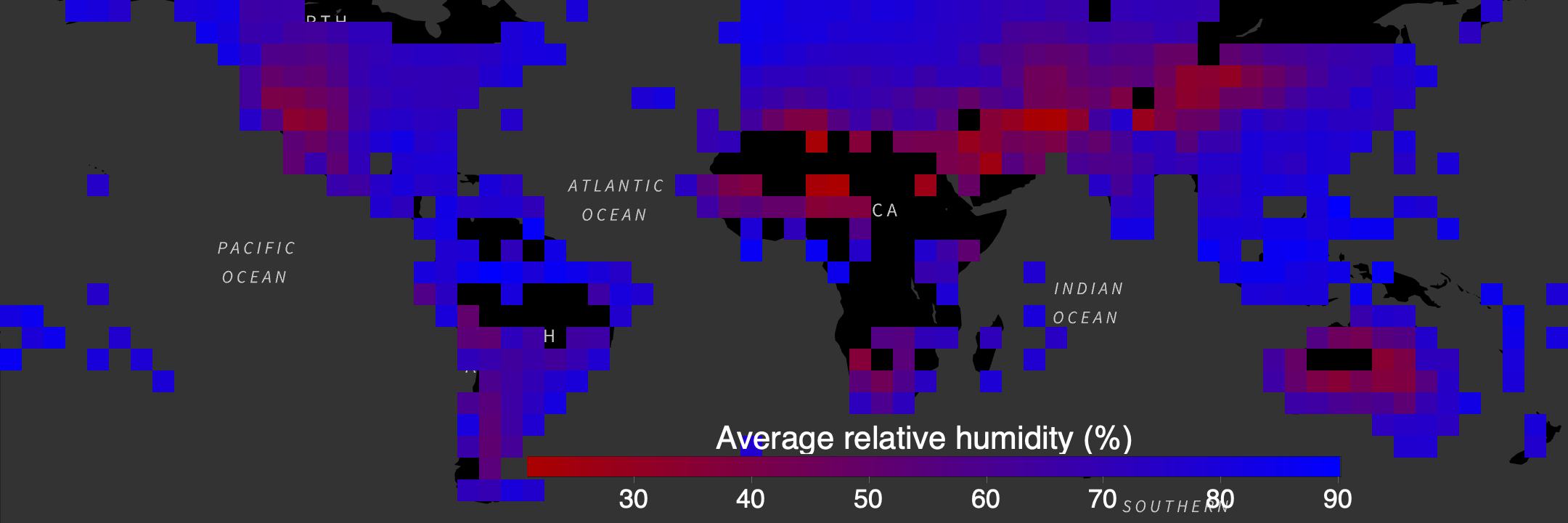}
    \caption{Global relative humidity data averaged over year 2021 from \cite{hadisdh}. Note that missing pixels are due to incomplete data in the original data set.}
    \label{fig:humidities}
\end{figure}

\begin{figure}
    \centering
    \includegraphics[width=\textwidth]{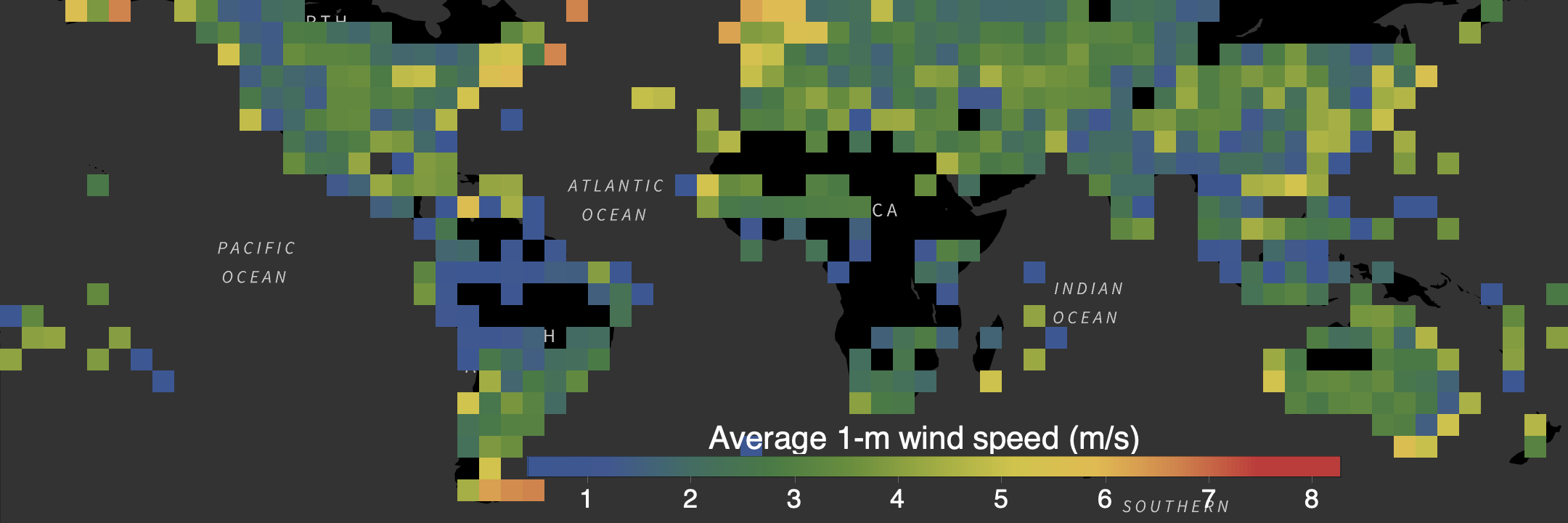}
    \caption{Global wind speeds adjusted to 1-m height from 10-m data \cite{gwa}.}
    \label{fig:windspeeds}
\end{figure} 

\begin{figure}
    \centering
    \includegraphics[width=\textwidth]{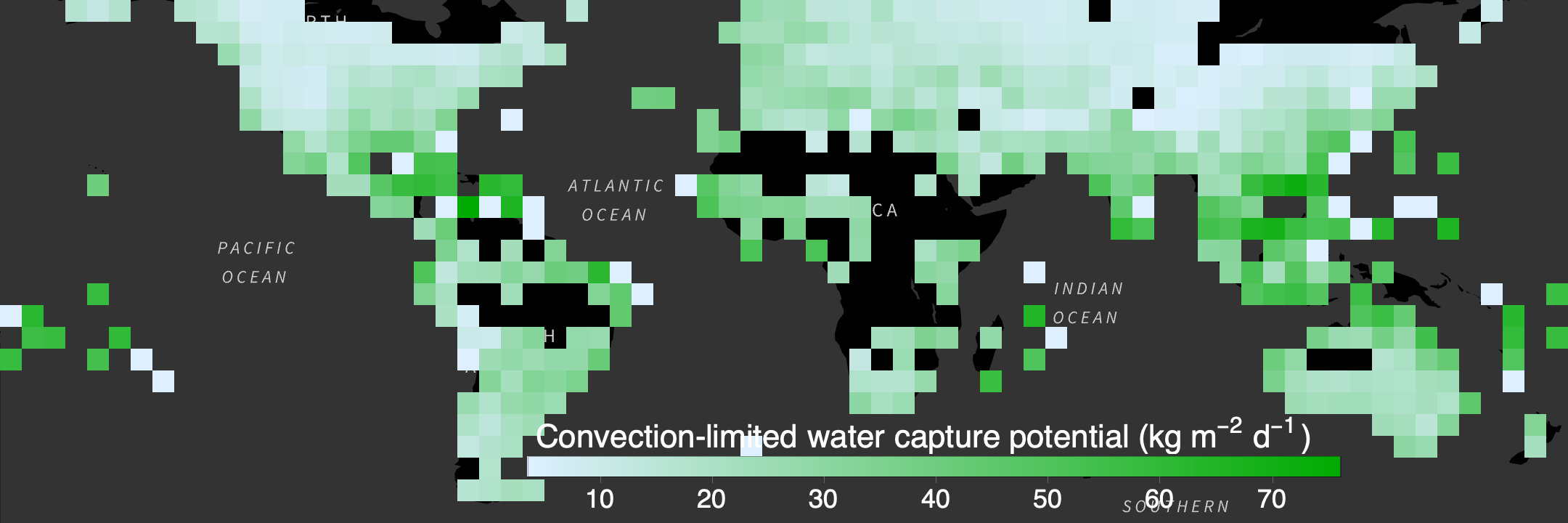}
    \caption{Potential convection-limtied water capture flux, $j$, calculated using Eq.~2 of the main text assuming device width of our prototype ($W=\SI{38}{\milli\meter}$).}
    \label{fig:capturepotential}
\end{figure}

\clearpage
\bibliography{sample}
